\documentclass[
 aip,
 amsmath,amssymb,
 reprint,
]{revtex4-1}

\usepackage[normalem]{ulem}
\usepackage{graphicx}
\usepackage{dcolumn}
\usepackage{bm}
\usepackage{verbatim,ragged2e,lipsum,pifont,subfigure}
\usepackage{amsmath,amsfonts,amssymb,amsbsy}
\usepackage{url}
\usepackage{upgreek}
\usepackage{mathrsfs}
\usepackage{euscript}
\usepackage{mathptmx}
\usepackage{xcolor}
\usepackage[utf8]{inputenc}
\usepackage[T1]{fontenc}

\newcommand{\rmd}{{\, \mathrm d}}
\newcommand{\rme}{{\, \mathrm e}}

\newcommand{\rmi}{{\mathrm i}}

\newcommand{\fkP}{{\mathfrak{P}}}
\newcommand{\fkY}{{\mathfrak{Y}}}

\newcommand{\RR}{{\mathbb R}}

\newcommand{\TT}{{\mathbb T}}

\begin{document}

\title[]{Low-dimensional chaos in the single wave model for self-consistent
wave-particle Hamiltonian}

 \author{J.~V.~Gomes}
 \email{janygovi@gmail.com}
 \affiliation{Universidade Federal do Paran\'a, Departamento de F\'isica, 81531-980, Curitiba, Paran\'a - Brazil}
 \affiliation{Aix-Marseille Universit\'e, CNRS, PIIM UMR 7345, 13397 Marseille - France}
 
 \author{M.~C.~de~Sousa}
 \email{meirielenso@gmail.com}
 \affiliation{Universidade de Sao Paulo, Instituto de Fisica, 05508-090, S\~ao Paulo, S\~ao Paulo - Brazil}
 \affiliation{Aix-Marseille Universit\'e, CNRS, PIIM UMR 7345, 13397 Marseille - France}

 \author{R.~L.~Viana}
 \email{viana@fisica.ufpr.br}
 \affiliation{Universidade Federal do Paran\'a, Departamento de F\'isica, 81531-980, Curitiba, Paran\'a - Brazil}

 \author{I.~L.~Caldas}
 \email{ibere@if.usp.br}
 \affiliation{Universidade de Sao Paulo, Instituto de Fisica, 05508-090, S\~ao Paulo, S\~ao Paulo - Brazil}
 
 \author{Y.~Elskens}
 \email{yves.elskens@univ-amu.fr}
 \affiliation{Aix-Marseille Universit\'e, CNRS, PIIM UMR 7345, 13397 Marseille - France}

\begin{abstract}
We analyze nonlinear aspects of the self-consistent wave-particle interaction using Hamiltonian dynamics in the single wave model, where the wave is modified due to the particle dynamics. This interaction plays an important role in the emergence of plasma instabilities and turbulence. 
The simplest case, where one particle $(N=1)$ is coupled with one wave $(M=1)$, 
is completely integrable,  
and the nonlinear effects reduce to the wave potential pulsating 
while the particle either remains trapped or circulates forever. 
On increasing the number of particles ($N=2$, $M=1$), integrability is lost and chaos develops. 
Our analyses identify the two standard ways for chaos to appear and grow 
(the homoclinic tangle born from a separatrix, and the resonance overlap near an elliptic fixed point). 
Moreover, a strong form of chaos occurs when the energy is high enough for the wave amplitude to vanish occasionally. 
\end{abstract}

\maketitle

\begin{quotation}
Wave-particle interaction plays an important role in plasma dynamics both in the laboratory and in space. 
The processes resulting from the interaction between charged particles and waves 
are related to the emergence of instability and turbulence in plasmas. 
In phase space, this interaction can generate both regular trajectories, 
which may lead to coherent particle acceleration, and chaotic trajectories, 
which are responsible for particle heating and escape. 
Low-dimensional approximations often shed light on the dynamics of systems with many degrees of freedom, 
as chaotic motion arises as one increases the number of degrees of freedom. 
In the simplest case, one particle $(N=1)$ is coupled to one wave $(M=1)$ in a self-consistent way, 
so that the wave is also modified due to the particle motion. 
This case is completely integrable, so that all trajectories are regular 
and the nonlinear effects degenerate to particle trapping or circulating while the wave potential pulsates. 
The bifurcation diagram of this simple system displays a saddle-center coalescence 
and a special trajectory for which the wave intensity goes through zero. 
On increasing the number of particles ($N=2$, $M=1$), chaos arises as this Hamiltonian system is not integrable. 
For low energy, chaos appears due to nonlinear resonances near the elliptic fixed point. 
For moderate energy, chaos appears and becomes more intense 
in the homoclinic tangle associated with the hyperbolic fixed points. 
For high enough energy, the wave phasor can pulsate through zero, 
and the sudden jump in its phase induces large-scale chaos. 
\end{quotation}

\section{\label{sec:Intro} Introduction}
Wave-particle interaction is one of the characteristic phenomena 
that occur naturally in plasma physics and play an essential role in their dynamics.\cite{ElEs03,EsEl03} 
Plasmas are naturally conducive to the amplification and propagation of waves due to their intrinsic tendency to restore balance in the local distribution of charges when the system is exposed to disturbances.\cite{Sw08} 
Attempts to make plasmas return to equilibrium can excite a diversity of wave modes, 
which are able to propagate in the plasma and interact with particles whose velocities are close to their phase velocity.\cite{St92} 
Hamiltonian systems provide a rich description of this interaction, 
where the regular and chaotic behavior of the particles trajectories in their phase space 
are directly related to the amplitude of the disturbance applied to the system. \cite{KaBe77,SmPe78}

The exchange of energy and momentum through wave-particle interaction is especially important in rarefied plasmas where the collision time between charged particles is generally very long compared to the characteristic time scales of the system, and therefore those plasmas can be treated as non-collisional. \cite{Balescu1988} At first, this implies that, in practice, there is no energy dissipation in low-density plasmas, since collisions are rare. However, the presence of waves can induce finite dissipation even in non-collisional plasmas:\cite{Ic18,El05} plasma particles are scattered by the wave fields, and their energies and momenta change through such processes.

In general terms, effective finite dissipation in collisionless plasmas occurs via resonance and can give rise, for example, to the growth/damping of waves and heating/acceleration of particles, as well as to the transport of charged particles.\cite{EsEl03,Besse2011,El12} The interaction becomes stronger when the streaming velocity of the particles is such that the particle couples with the Doppler-shifted wave at its cyclotron frequency or its harmonics. This is the so-called cyclotron resonance interaction.\cite{Ti74} The special case of the Doppler-shifted wave frequency being zero ({\sl i.e.} zero harmonic of the cyclotron frequency) corresponds to the well-known Landau resonance.\cite{EBEZD18}

In practice, Landau damping (resp.\ growth) can be understood as follows: as observed experimentally,\cite{DoEsMa05} particles with velocities slightly lower (larger) than the phase velocity of a wave are accelerated (decelerated) by the wave's electric field. Thus, particles that move a little slower (faster) than the phase velocity gain (lose) energy 
from (to) the wave. \cite{Chen1984}

The main concepts described by Landau are widely used in particle accelerators to avoid instabilities in the coherent oscillation of the beams.\cite{Herr13} Besides, aspects of this interaction are notoriously important in space plasma physics, such as in the suprathermal electron acceleration at the solar wind, \cite{HWTMZ15} in the interaction of charged particles with the Earth's magnetic field,\cite{ChKlHo19} etc. For this reason, even many decades after its discovery, \cite{La46} there is high interest in the fundamental aspects related to Landau damping. \cite{Ryutov99,StSu99}

An important feature of this type of interaction is that ions, being much more massive than electrons, are assumed to be fixed and their role is limited to providing charge neutrality for the system. The collective vibration of electrons with respect to ions is called Langmuir waves. \cite{St92} The usual description of the interaction of Langmuir waves with electrons whose velocities are close to their phase velocity involves the kinetic set of Vlasov-Poisson equations for the electron distribution function. \cite{ElEsDo14}

In order to describe the interaction between charged particles and electrostatic waves, it is natural to use Hamiltonian models for which the particle dynamics in phase space generates both regular and chaotic trajectories. \cite{KaBe77,SmPe78} The predominance of either type of trajectory depends mainly on the amplitude of the perturbation in the system that directly influences the particles motion. \cite{Es85, MFRF10} In general, regions where regular trajectories prevail are more favorable to coherent particle acceleration, while chaotic regions are associated with particle heating and escape. \cite{IKK83}

Wave-particle interactions have often been described by Hamiltonian models in which particle motion is affected by the wave field, whereas the wave itself is not influenced by particle motion. \cite{KaBe77,SmPe78} However, proper treatment of the problem would require also the addition of the wave response to the particle motion, which leads to so-called self-consistent Hamiltonians. \cite{MyKa78,ElEs03} In this framework, the dynamics of Langmuir waves is described as $M$ harmonic oscillators coupled to $N$ quasi-resonant particles. Considering the single wave model (SWM) introduced by Onishchenko, O'Neil, and coworkers, \cite{OLM70,OWM71} it is possible to study the chaotic dynamics of wave-particle self-consistent interaction in terms of a few degrees of freedom. Indeed, this model can even be reduced to a four degrees of freedom system to describe its saturation regime, \cite{TenMeMo94,Antoniazzi2006} and the model with a single particle was already considered by Adam, Laval and Mendon\c{c}a \cite{Adam1981} with a view at its integrability and at the generation of sideband modes of the Langmuir waves. 

The SWM originates from the description of the beam-plasma instability 
and has the advantage of behaving smoothly when the number of particles tends to infinity.\cite{TenMeMo94,FiEl98}  
Since its introduction, this model has proven to be relevant in a variety of physical situations in which the dynamics is effectively dominated by a single mode as in the confinement of charged particles in tokamaks, \cite{Carlevaro2016} Landau damping, \cite{FiEl00,YaFi09} free-electron lasers, \cite{HuKi07,Antoniazzi2006} in the relationship between self-consistent chaos and phase space coherent structures, \cite{CaFi02} and in kinetic instabilities of the Alfv\'en wave-particle interaction obtained experimentally in tokamak JET. \cite{TesFas04} 

In the present work, we revisit the dynamics of the single wave model with one particle $(N = 1)$, which is integrable, so that the phase portrait comprises only regular trajectories. The bifurcation diagram, in this case, shows a saddle-center coalescence that occurs for a specific value of total momentum $P$ and divides the phase portrait topologies. Moreover, we stress the role of the trajectory for which the wave intensity $I$ passes through zero, and we find a specific value of total momentum for which this trajectory coincides with a branch of the separatrix. 

For two particles $(N = 2)$, we study the emergence of low-dimensional chaos. 
We observe that the intensification of chaotic activity occurs 
both in the domains close to the elliptic fixed point and close to the separatrix associated with a hyperbolic fixed point. 
Fourier analysis shows that the nonlinear evolution of the particles motion, close to the elliptic fixed point, 
gives rise to the appearance and intensification of resonances. 
At higher energy, for which a hyperbolic point exists, the system is significantly chaotic. 
Moreover, for a still larger energy, the wave intensity can pass through zero, 
and the system exhibits chaos on a larger scale. 

Our numerical computations were performed using a leap-frog symplectic integrator, which conserves the geometry of the system exactly and its energy quite accurately for long time.\cite{HLW06} For the non-integrable case with two particles, we studied the dynamics by intercepting the trajectories with a Poincar\'e section. \cite{Zas05}

This article is organized as follows: in the next section (\ref{sec:PhySet}), we present the single wave Hamiltonian. 
The dynamics for one particle $(N=1)$ is revisited in section~\ref{sec:M1N1}. 
General aspects of the two-particle model are discussed in section~\ref{sec:M1N2}. 
Fixed points and special trajectories are analyzed in section~\ref{sec:FixPt_Traj_N2}, 
while section~\ref{sec:TrajectoriesM1N2} presents Poincar\'e sections and the time evolution of typical trajectories. 
The last section (\ref{sec:Conclusion}) is devoted to our conclusions and prospects.

\section{The single wave Hamiltonian}
\label{sec:PhySet}
The self-consistent dynamics of $N$ identical particles moving on the interval of length $L$ with periodic boundary conditions, interacting with $M$ longitudinal waves with wave numbers $k_j=j 2\pi/L$ and natural frequencies $\omega_{0j}$, is described by the reference Hamiltonian~\cite{ElEs03,El05}
\begin{eqnarray}
  H_{\rm sc}^{N,M} &=& \sum_{r=1}^N {\frac {p_r^2} {2 m_r}} \ +\ \sum_{j=1}^M \omega_{0j} {\frac {X_j^2 + Y_j^2} 2} \ +\ \nonumber  \\  && 
  + \varepsilon \sum_{r=1}^N \sum_{j=1}^M k_j^{-1} \beta_j (Y_j \sin k_j x_r - X_j \cos k_j x_r),
  \label{eqHXY}
\end{eqnarray}
\begin{eqnarray}
  H_{\rm sc}^{N,M} &=& \sum_{r=1}^N {\frac {p_r^2} {2 m_r}} \ +\ \sum_{j=1}^M \omega_{0j} I_j \ + \ \nonumber  \\ &&  
  - \varepsilon \sum_{r=1}^N \sum_{j=1}^M k_j^{-1} \beta_j \sqrt{2 I_j} \cos (k_j x_r - \theta_j),
  \label{eqHItheta}
\end{eqnarray}
where $\beta_j $ is the coupling constant of wave $j$ and $\varepsilon$  is the overall coupling parameter. Here, $Z_j = X_j + \rmi Y_j = \sqrt{2 I_j} \rme^{- \rmi \theta_j}$, the generalized coordinates are the particles positions $x_r$ and waves phases $\theta_j$, and their conjugate momenta are the particles momenta $p_r$ and waves intensities $I_j$. In phasor formulation, wave $j$ has $X_j$ as generalized coordinate with conjugate momentum $Y_j$. 

Hamiltonian $H_{\rm sc}^{N,M}$ comprises three contributions: the free motion (kinetic energy) of the particles, the (harmonic) oscillation of the waves, and the coupling between particles and waves. Besides that, Hamiltonian $H_{\rm sc}^{N,M}$ is invariant under translation in time and in space so that the total energy $E = H_{\rm sc}^{N,M}$ and the total momentum $P =  \sum_{r=1}^N p_r +  \sum_{j=1}^M k_j I_j$ are conserved. The latter constant reveals that the growth or decay of a wave is directly balanced with the slowing down or acceleration of particles.

We focus on a single special case, where all particles have the same mass, and we rescale time and energy to set the coupling constant $\varepsilon \beta_1$ and the particles mass $m$ equal to unity in Eq.~(\ref{eqHSWM}). In the single wave model ($M=1$), we omit the subscript $j$ and set the length unit to $k^{-1}$ and the spatial period to $L=2\pi$, which reduces the Hamiltonian to
\begin{eqnarray}
  H_{\rm sc}^{N} 
  = \sum_{r=1}^N {\frac {p_r^2} {2}} \ +\ \omega_{0} I \ -\ \sqrt{2 I} \sum_{r=1}^N  \cos ( x_r - \theta).
  \label{eqHSWM}
\end{eqnarray}

A Galileo transformation enables us to put the system in the reference frame of the wave. With the generating function $F_1(x,\theta,\bar{p},\bar{I},t)=\sum_{r=1}^N(x_r - \omega_0 t) (\bar{p}_r + \omega_0) + (\theta -\omega_0 t)\bar{I} - N  \omega_0^2 \, t /2 $, Hamiltonian~(\ref{eqHSWM}) becomes
\begin{eqnarray}
  \bar{H}(\bar{p},\bar{I}, \bar{x}, \bar{\theta}) 
  &=& H_{\rm sc}^{N} + \frac{\partial F_1}{\partial t}, \nonumber  \\
  &=& \sum_{r=1}^N {\frac {\bar{p}_r^2} {2}} \ -\  \sqrt{2 \bar{I}} \sum_{r=1}^N  \cos ( \bar{x}_r - \bar{\theta}). 
  \label{eqHswm2}
\end{eqnarray}

Total momentum, 
\begin{eqnarray}
\bar{P} =  \sum_{r=1}^N \bar{p}_r + \bar{I},
\end{eqnarray}
is conserved by the dynamics obtained from Eq.~(\ref{eqHswm2}). This enables us to define a new generating function  $F_2(\bar{x}, \bar{\theta}, p', I') = I' \bar{\theta} + \sum_{r=1}^N p'_r (\bar{x}_r - \bar{\theta})$: the new coordinate conjugate to $p'_r = \bar{p}_r$ is $x'_r = \partial F_2/\partial p'_r = (\bar{x_r} - \bar{\theta})$, which we denote as $y_r = x'_r$, and the new momentum conjugate to $\theta' = \bar{\theta}$ is $I' = \bar P$. The latter is a constant of motion so that the new angle $\theta' = \bar{\theta}$ is a cyclic coordinate. The final Hamiltonian, emphasizing that only $N$ degrees of freedom are effective, is obtained in the compact form
\begin{eqnarray}
  H(p,y) =  \sum_{r=1}^N {\frac {p_r^2} {2}} \ -\  \sqrt{2 I} \sum_{r=1}^N  \cos y_r , 
  \label{eqHFswm}
\end{eqnarray}
where, for short, we dropped the prime from $p'_r$ and the overbars from $\bar{I} = \bar{P} - \sum_r p'_r$ and from $\bar{H}$. 

Wave-particle interaction is typical in many physical systems, and we investigate in this paper how this particular form of coupling given by Hamiltonian~(\ref{eqHFswm}) affects the particles dynamics as we increase the number of degrees of freedom. 
This single wave Hamiltonian was first formulated as a simplified model to treat the instability due to a weak cold electron beam in a plasma, assuming a fixed ionic neutralizing background.\cite{OLM70,OWM71} More recently, different studies extended the application of the single wave model to a much larger class of instabilities,\cite{CrJa99} derived it in a generic manner from different contexts, and proved it could model various phenomena in fluids and plasmas,\cite{CaFi02} and Compton free-electron laser amplification.\cite{FCCP94}

\section{The single wave with one particle}
\label{sec:M1N1}
In order to understand this system, we start with a few degrees of freedom. 
Following Adam, Laval, Mendon\c{c}a, Tennyson, Meiss, Morrison and del-Castillo-Negrete 
and recalling results from Refs~\onlinecite{TenMeMo94,Adam1981,Ca02}, 
we first study the simplest, integrable case for this model 
where the self-consistency couples one particle and one wave, $M=N=1$. 
As we will see in section~\ref{sec:M1N2}, the dynamics for $N = 2$ incorporates 
most of the basic phenomena that we will discuss in the next two subsections. 
Moreover, the dynamics with $N = 1$ bears fundamental importance 
in the description of phenomena for the case with many particles, \cite{TenMeMo94,Adam1981} 
where the macroparticle is used to describe the dynamics of an electron beam 
so that the beam electrons oscillate bunched at the bottom of the wave potential well 
during the trapping process.\cite{OLM70,OWM71,TenMeMo94,Antoniazzi2006}

\subsection{Preliminary analysis of the dynamics for $N = M = 1$}
\label{sec:PrelAnalM1N1}
The single wave Hamiltonian for this case reads
\begin{eqnarray}
  H 
  = \frac{p^2}2 + (Y \sin x - X \cos x)
  = \frac{p^2}{2} - \sqrt{2 I} \cos (x - \theta),
\label{eqH11}
\end{eqnarray}
and the conserved total momentum is $P = p + I$. 
The evolution equations
\begin{subequations}
\begin{eqnarray}
  \dot x & = & p,
  \label{eq:dotx1}
  \\
  \dot p & = & - X \sin x - Y \cos x = - \sqrt{2 I} \sin (x - \theta),
  \label{eq:dotp1}
  \\
  \dot X & = & \sin x,
  \label{eq:dotX1}
  \\
  \dot Y & = & \cos x,
  \label{eq:dotY1}
\end{eqnarray}
\end{subequations}
imply that ${\dot X}^2 + {\dot Y}^2 = 1$ so that the wave never remains still. Besides, $\ddot p = - 1 + (Y \sin x - X \cos x) p$. 

For this simple case with only one particle, the single wave Hamiltonian has two degrees of freedom, one for the particle and one for the wave. As the Hamiltonian is invariant under space translations, the momentum conservation law reduces the problem to one degree of freedom. To express this, we introduce $y = x - \theta$ and write the Hamiltonian in the form 
\begin{equation}
  H = \frac{p^2}{2} - \sqrt{2 (P - p)} \, \cos y,
\label{eqH11yv}
\end{equation}
with $I$ expressed in terms of the particle momentum. As this Hamiltonian is time-independent, the system is completely integrable. In particular, particle orbits in phase portrait follow the constant energy contours $(H=$ constant). 

The equations of motion of Hamiltonian~(\ref{eqH11yv}) read
\begin{subequations}
\begin{eqnarray}
  \dot y & = & p + \frac{1}{\sqrt{2 (P - p)}} \cos y,
  \label{eq:doty}
  \\
  \dot p & = &  - \sqrt{2 (P - p)} \sin y.
  \label{eq:dotv}
\end{eqnarray}
\end{subequations}
The fixed points of the system are defined by the conditions 
\begin{eqnarray}
  \dot y = \partial_p H = 0, \hspace{1cm} \dot p = - \partial_y H = 0.
  \label{eq:fpc}
\end{eqnarray}
Solving these conditions for Hamiltonian~(\ref{eqH11yv}), we obtain the coordinates  $(y_i^*,p_i^*)$ of the fixed points $C_i^*$
\begin{subequations}\label{fpy}
\begin{eqnarray}
  C_1^* &:& ( 0 \, , \, p_1^* \sqrt{2(P-p_1^*)}=-1), 
  \label{fpy0} \\
  C_{2,3}^* &:& ( \pi \, , \, p_{2,3}^* \sqrt{2(P-p_{2,3}^*)}=1), 
  \label{fpypi}
\end{eqnarray}
\end{subequations}
with $p_1^* < 0$, $0 < p_2^* < 1$, $p_3^* > 1$, and the wave intensity at the fixed points given by $I_i^* = P - p_i^*$.

The stability of $C^*_i$ 
is determined from the eigenvalues $\lambda_i$ of the Jacobian matrix by linearizing the equations of motion (\ref{eq:doty}) and (\ref{eq:dotv}) in the vicinity of each fixed point. Doing so, we find that the eigenvalues for $C^*_1$ and $C^*_{2,3}$ are respectively 
\begin{subequations}
\begin{eqnarray}
  \lambda_1 & = & \pm \left( -\sqrt{2 I_1^*} - \frac{1}{2 I_1^*} \right)^{1/2},
  \label{eigL1}
  \\
  \lambda_{2,3} & = & \pm \left( \sqrt{2 I_{2,3}^*} - \frac{1}{2 I_{2,3}^*} \right)^{1/2}.
  \label{eigL2}
\end{eqnarray}
\end{subequations}

The eigenvalue $\lambda_1$ is imaginary for any value of $I^*_1$, which means that the fixed point $y^*_1=0$ has elliptic stability. In addition, since $p_1^* < 0$, the physical condition $I_1^* = P - p_1^*$ implies that $I_1^* > P$ for any value of $P$.

When $ I^*_{2} = I^*_{3} = 1/2 $, the eigenvalues $\lambda_{2,3}=0$, indicating a bifurcation point at $y_{2,3}^*=\pi$ that occurs for $P=3/2$ and $p_2^*=p_3^*=1$. 
For $ I^*_2 > 1/2 $, the eigenvalue $\lambda_2$ is real, so that in $y^*_2=\pi$ the system has hyperbolic stability for any $P>3/2$ with $0<p_2^*<1$. Finally, for $ 0 < I^*_3 < 1/2 $, the eigenvalue $ \lambda_3 $ is imaginary, indicating that at the same abscissa $y^*_3 =y^*_2 =\pi$ we also have elliptic stability for any $P>3/2$ with $p_3^*>1$.

\begin{figure}[!tb]
\centering
   \subfigure[]{
     \includegraphics[width=0.45\textwidth]{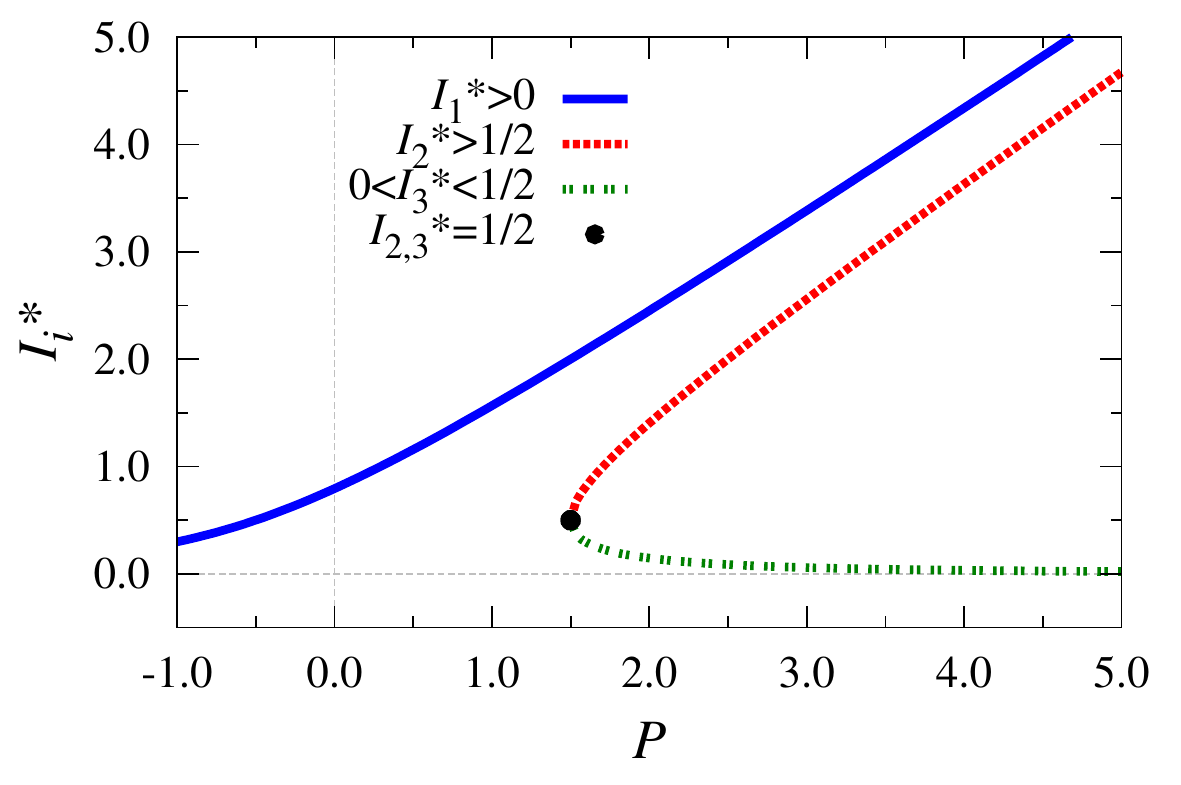}
     \label{01a} }
   \subfigure[]{
     \includegraphics[width=0.45\textwidth]{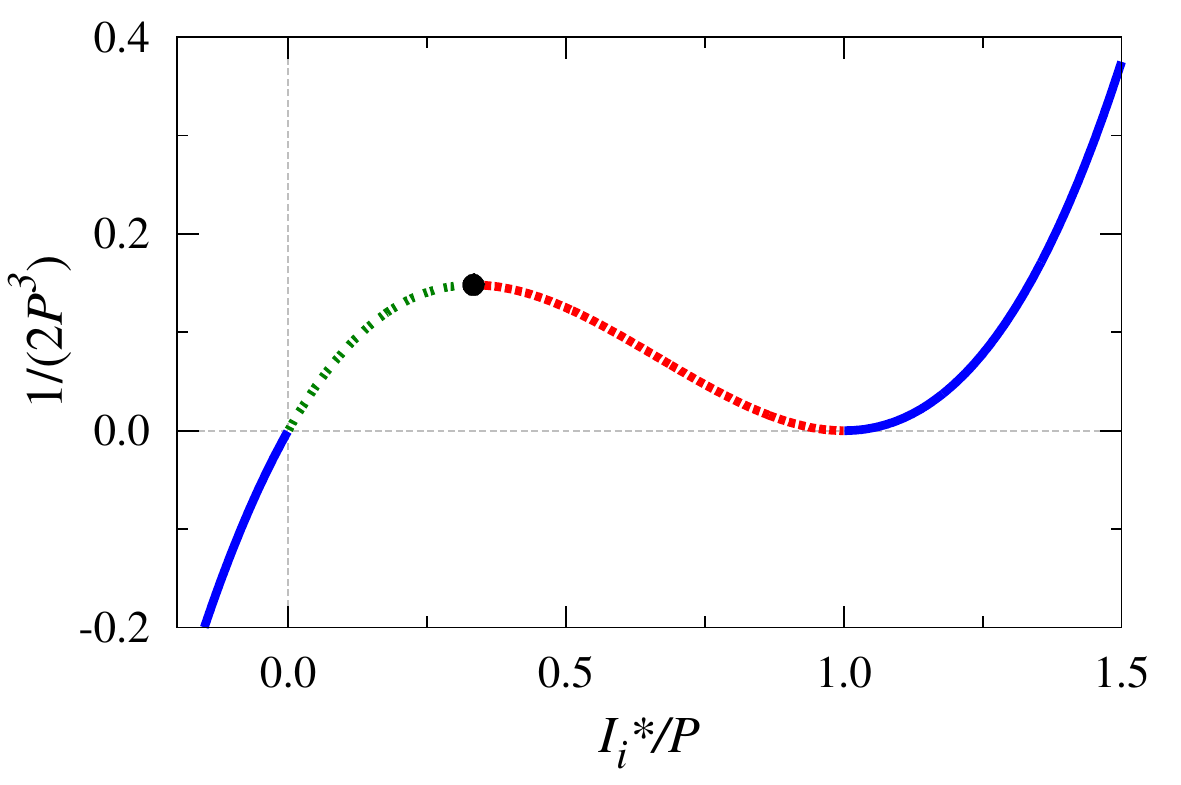}
     \label{01b} }
   \caption{(a) Bifurcation diagram of equation (\ref{eq:BifEq}) for the $M = N = 1$ system. The blue (solid) line corresponds to the elliptic stable fixed point at $ y_1^* = 0 $. The black point at $ I_{2,3}^* = 1/2 $ and $ P = 3/2 $ corresponds to the bifurcation, and the red (dotted) and green (double-dotted) lines correspond, respectively, to the hyperbolic and the elliptic fixed points at $ y_{2,3}^* = \pi $ after bifurcation. (b) Roots of the normalized equation (\ref{eq:BifEq}), $( 1 - I_i^*/P )^2 \, I_i^*/P = 1 / (2 P^3)$.}
  \label{fig:BifDiag}
\end{figure}

The values of $p_i^*$ at the fixed points (\ref{fpy}) are obtained as a function of total momentum $P$, i.e. $p_i^* = \pm 1/\sqrt{2(P-p_i^*)}$,  so that we can describe the equilibrium solutions with the equation
\begin{equation}
  (P-I_i^*)^2 I_i^* = 1/2.
  \label{eq:BifEq}
\end{equation}
As shown in Fig.~\ref{fig:BifDiag}, equation (\ref{eq:BifEq}) selects the $I_i^*$ values for which the cubic polynomial on the left-hand side assumes a given value. The blue (solid) line represents the stable solution at the elliptic fixed point at $y_1^* = 0$ : this solution exists for any value of total momentum $P$. The black point at $P=3/2$ shows a bifurcation, where two types of equilibrium with different stability coincide at the same fixed position $y_{2,3}^* = \pi$. After the bifurcation point, the red (dotted) line corresponds to the unstable solution at the fixed position $y_2^* = \pi$, and the green (double-dotted) line corresponds to the stable solution in the same fixed position $y_3^* = \pi$.

The $M = N = 1$ system is integrable. Actually, solving (\ref{eqH11yv}) for $\cos y$ and squaring (\ref{eq:dotp1}) leads to the first order equation 
\begin{eqnarray}
   \dot p^2 
   &=& 2 (P - p) \left[ 1 - \left(\frac{H - p^2 / 2}{\sqrt{2 (P - p)}}\right)^2 \right],
   \nonumber \\
   &=&  2 P - H^2 - 2 p + H p^2 - \frac{p^4}{4} ,
   \label{eq:dotpell}
\end{eqnarray}
which is solved analytically in terms of elliptic functions.\cite{Adam1981} 
Briefly, one finds a function $\fkP$ such that $p = \fkP(t ; p^*, P, H)$ by integrating (\ref{eq:dotpell}), and a function $\fkY$ such that $y = \fkY(t ; p^*, P, H) = {\mathrm{Arcsin}} [ - \dot p / \sqrt{2(P-p)} ] 
= {\mathrm{Arccos}} [ (p^2 - 2 H)/\sqrt{8 (P - p)} ]$ modulo boundary conditions. 
One can also construct action-angle variables for each type of periodic trajectory. 

The equilibrium points $p={p_i^*}=$~constant for equation~(\ref{eq:dotpell}) are defined by the conditions
\begin{eqnarray}
   G(p_i^*) = 0, \hspace{0.5cm} \frac{\rmd G}{\rmd p}\Bigg|_{p=p_i^*} = 0, 
   \label{eqpdG}
\end{eqnarray}
with $G(p)$ the quartic polynomial on the right-hand side of (\ref{eq:dotpell}). 

Solving (\ref{eqpdG}) for the values of parameters $H$ and $P$, 
we find parametrically given curves
\begin{eqnarray}
    P({p_i^*}) = \frac{1 + 2 {p_i^*}^3}{2 {p_i^*}^2}, \hspace{0.5cm} 
    H({p_i^*}) = \frac{2 + {p_i^*}^3}{2 {p_i^*}}. \label{PCPH}
\end{eqnarray}
These curves on the $(P, H)$ plane contain important information on the system dynamics. \cite{bolsinov2010} 

\begin{figure}[!tb]
\centering
   \includegraphics[width=0.45\textwidth]{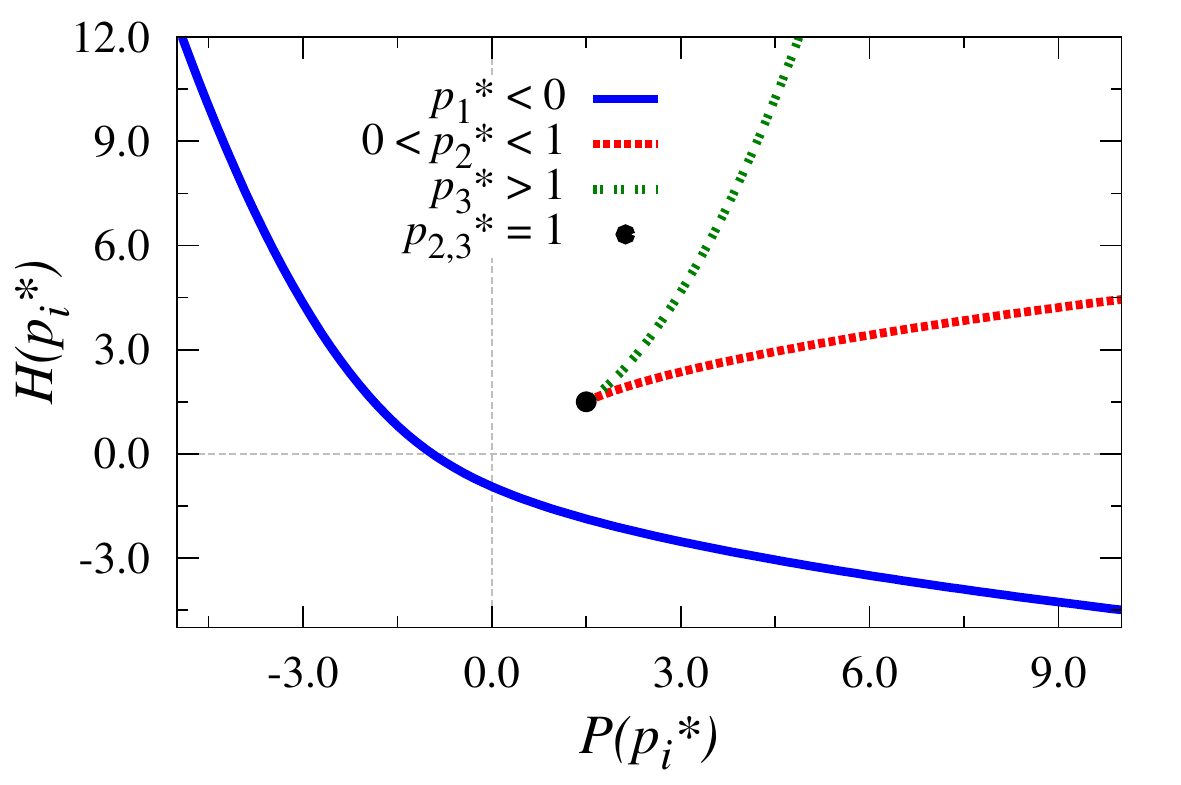}
   \caption{Curves on the $(P,H)$ plane for equations (\ref{PCPH}).}
  \label{fig:PxH}
\end{figure}

The loci of equations (\ref{PCPH}) on the $(P,H)$ plane are shown in Fig.~\ref{fig:PxH}.
As in Fig.~\ref{01a}, the blue (solid) curve 
represents the stable elliptic point at $y_1^* = 0$; 
the black point at $(P, H) = (3 / 2,3 / 2)$ with ${p_{2,3}^*} = 1$ corresponds to bifurcation; 
the red (dotted) curve represents the parameters of the hyperbolic fixed point at $y_2^* = \pi$; 
and the green (double-dotted) line is associated with the elliptic fixed point also at $y_3^* = \pi$. 
After bifurcation, for a given value of $P$, the energy of the stable (elliptic) fixed point at $y_1^* = 0$ is lower than the energy of the hyperbolic fixed point at $y_2^* = \pi$, 
and the latter, in turn, is lower than the energy of the elliptic fixed point at $y_3^* = \pi$. 
The topological changes described by the solutions of (\ref{eq:BifEq}) in Fig.~\ref{01a} and by (\ref{PCPH}) in Fig.~\ref{fig:PxH} are presented in the phase portraits of Section \ref{sec:PhasePortraitM1N1}.

Moreover, for $(P, H)$ on these curves, the evolution equation (\ref{eq:dotpell}) reduces to
\begin{eqnarray}
   \dot{p} = \pm (p - {p^*}) \Bigg( \frac{1}{{p^*}} - \frac{(p - {p^*})^2}{4} \Bigg)^{1/2} \, ,
   \label{eq:separatrix}
\end{eqnarray}
which can be solved in terms of elementary functions. Specifically, if $0 < p^* < 1$, this equation admits real-valued solutions for real time, describing motion on the separatrix of the hyperbolic fixed point. On the contrary, if $p^* < 0$ or if $p^* > 1$, equation (\ref{eq:separatrix}) has no real-valued solution as the associated fixed point is elliptic.

\subsection{Phase portrait analysis for $N = M = 1$}
\label{sec:PhasePortraitM1N1}
\begin{figure*}[]
  \centering
   \subfigure[]{\includegraphics[width=0.42\textwidth]{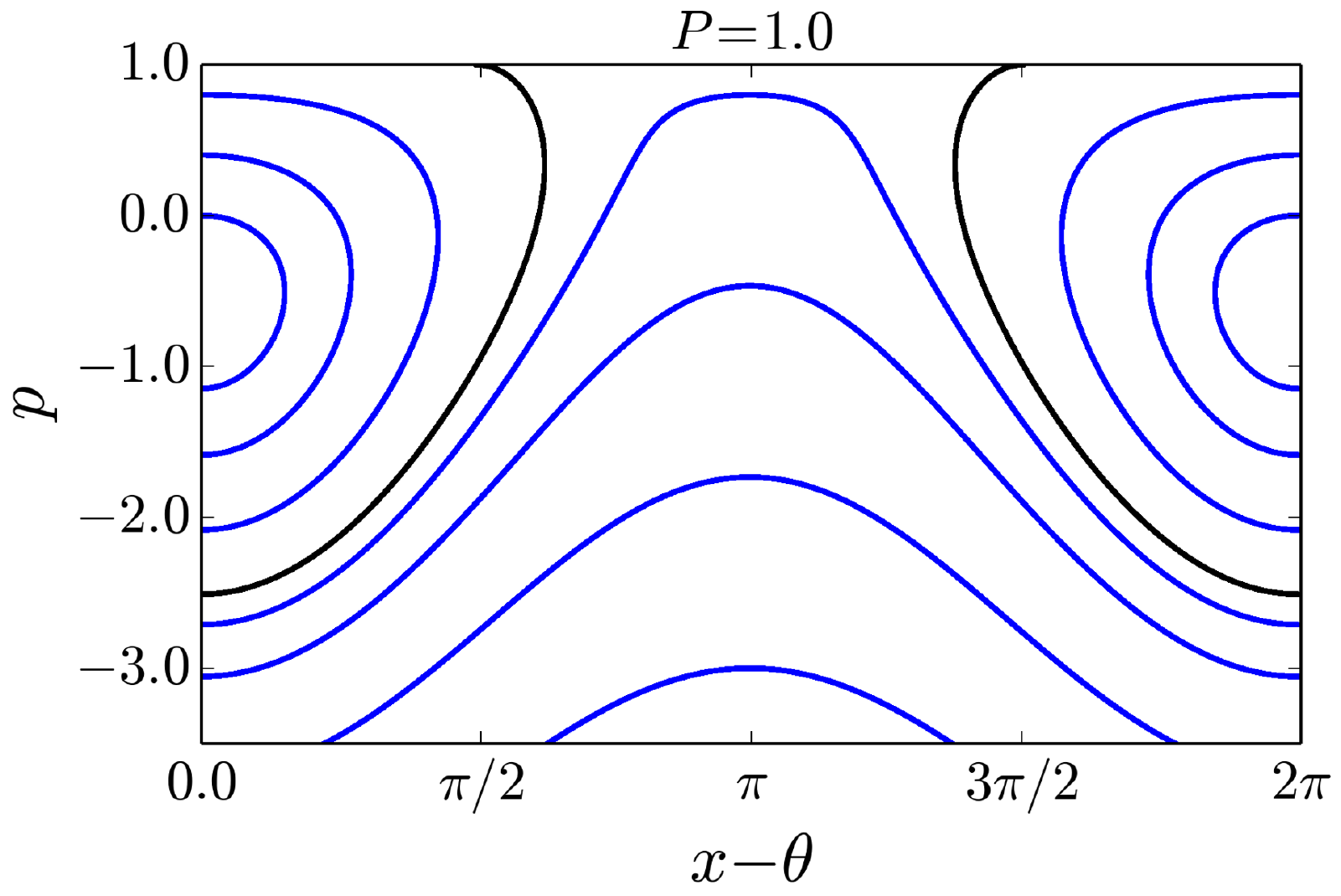}\label{fig:03a} }\hspace{0.5cm}
   \subfigure[]{\includegraphics[width=0.42\textwidth]{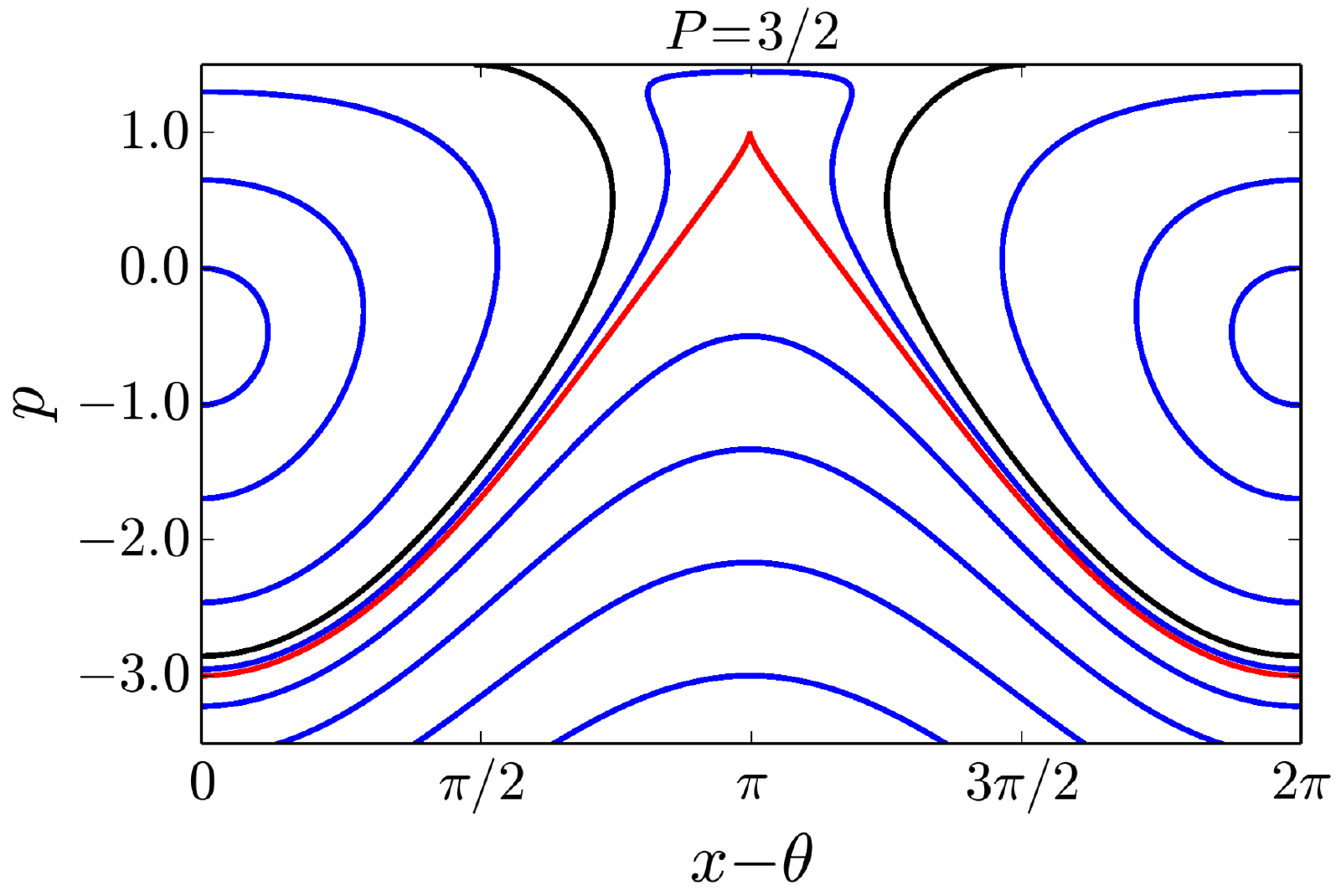}\label{fig:03b} }\\
   \subfigure[]{\includegraphics[width=0.42\textwidth]{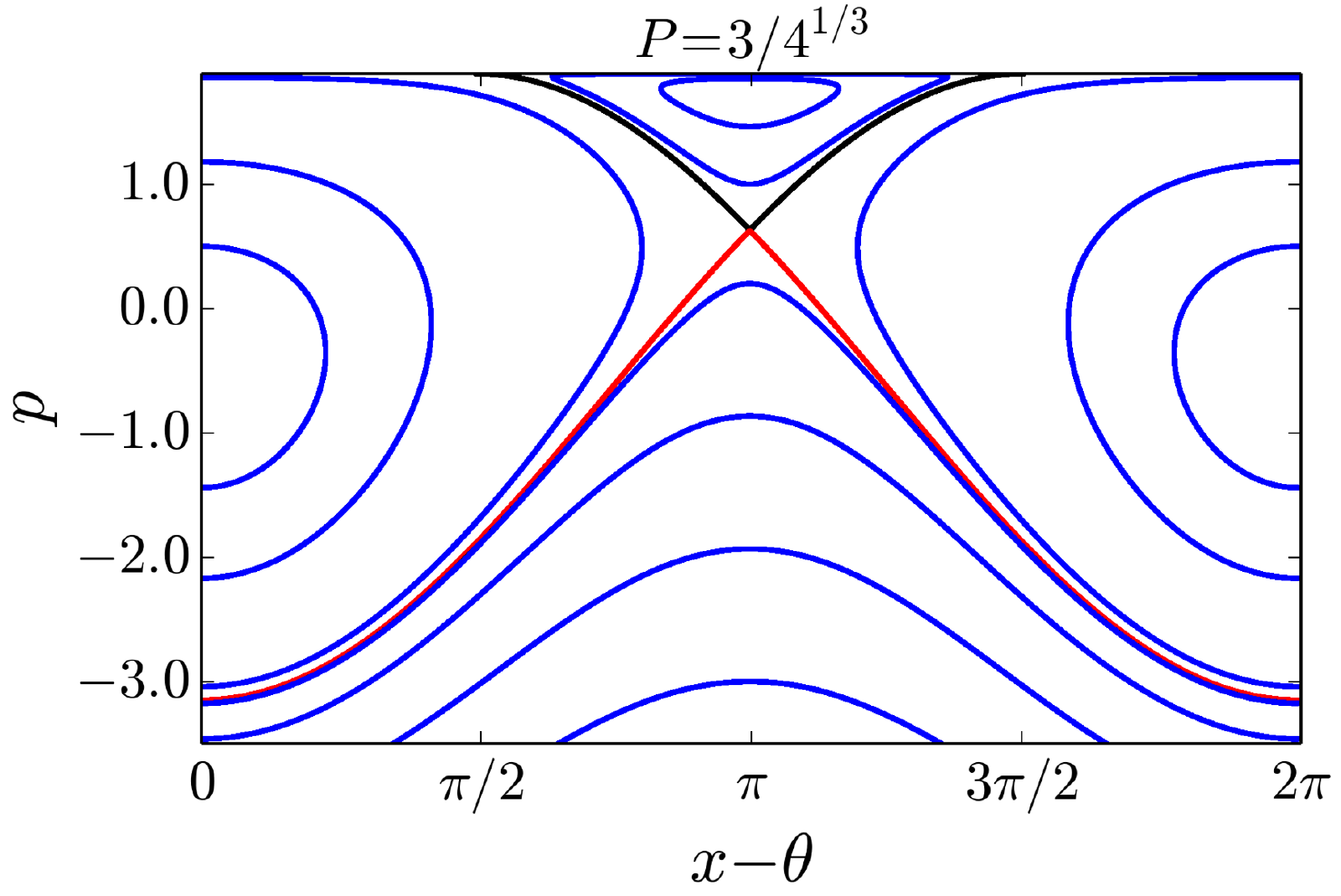}\label{fig:03c} }\hspace{0.5cm}
   \subfigure[]{\includegraphics[width=0.42\textwidth]{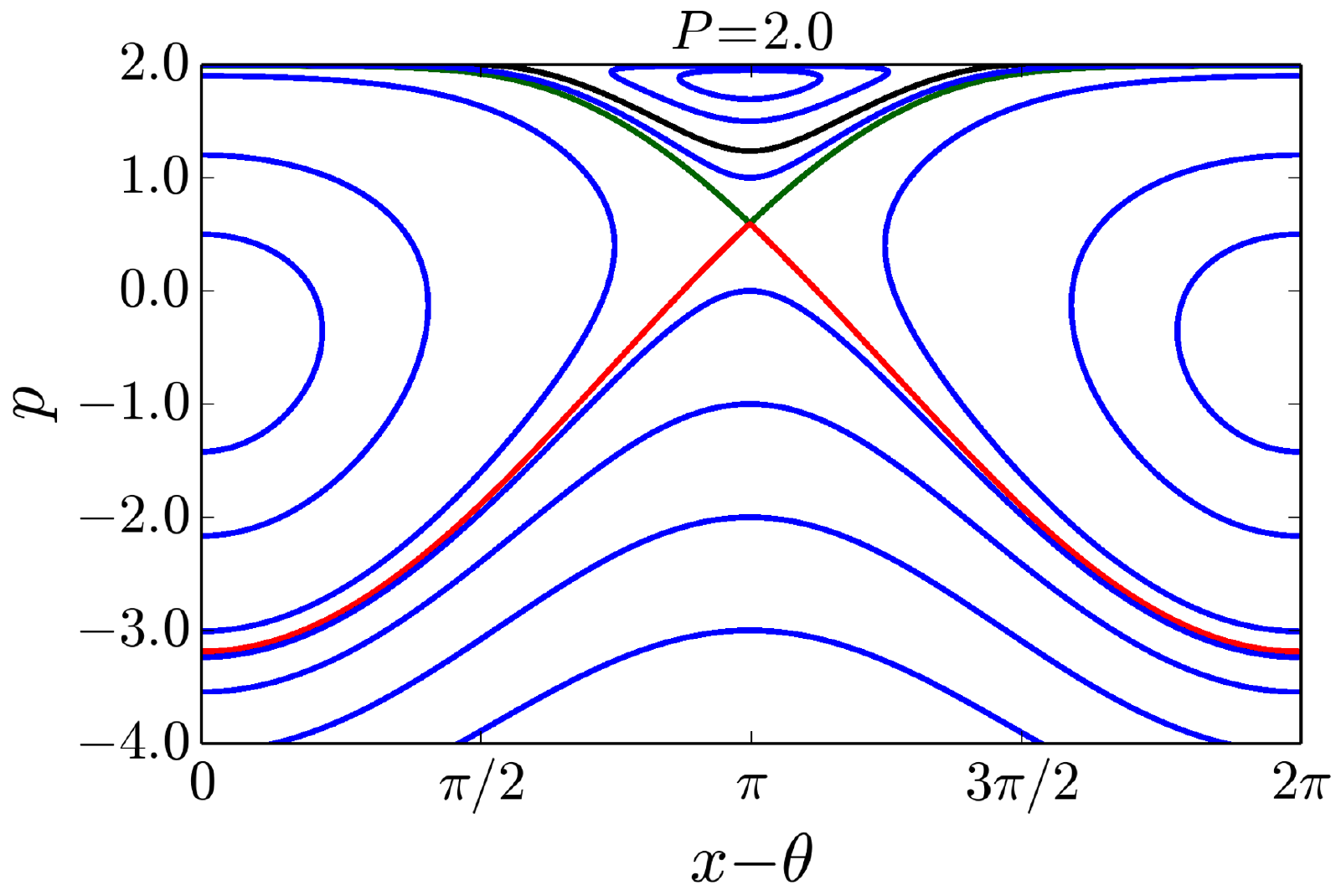}\label{fig:03d} }
   \caption{Phase portrait with $N=1$ for the single wave Hamiltonian~(\ref{eqH11yv}). Panel (a) shows the case $P<3/2$ (before bifurcation), panel (b) shows the case $P=3/2$ (at the saddle-center bifurcation), panel (c) corresponds to total momentum $P=3/4^{1/3}$, for which the trajectory containing $I=0$ coincides with the upper branch of the X point separatrix, and panel (d) shows the dynamics after the global bifurcation.}
   \label{fig:03}
\end{figure*}

The phase portrait of the system in the $(p, y)$ variables is shown in Figure~\ref{fig:03} and has special boundaries. Indeed, variable $y=x-\theta$ is $2\pi$-periodic and the wave intensity must be positive so that $p \leq P$, and the portrait will be plotted over half a cylinder. 

As already seen in the previous subsection, the fixed point at $y_1^* = 0$ has elliptic stability for all values of total momentum $P$, so that the system dynamics around the elliptic point is represented by closed trajectories.

The black line in the phase portraits represents the trajectory for which the wave intensity $I$ passes through $0$. 
The ordinate $P$ for $p$ does not correspond to a continuum of values for $y$, because equation (\ref{eq:doty}) is meaningless if $\cos y \neq 0$. 
Thus only abscissae $y = \pm \pi/2$ are permitted when $I = 0$, and then the wave phase is actually undefined. But the dynamics is well-defined in cartesian variables $(X, Y)$ and, if the wave turns out to vanish at a time, then ${\dot X}^2 + {\dot Y}^2 = 1$ implies that $I$ cannot remain zero, {\sl i.e.}, the potential acting on the particle cannot remain flat. Actually, as the particle position is a smooth function of time, what occurs when $I$ vanishes is that the wave phase jumps between $x - \pi/2$ and $x + \pi/2$, and the trajectory in $(p, y)$ variables transits through this connection with $\dot p = 0$ and $\ddot p = -1$, so that the value $p=P$ is a non-degenerate local maximum of $p$ along the trajectory. According to (\ref{eqH11yv}), this trajectory has energy $H = P^2 / 2$. The phase jump by $\pi$ instantly interchanges the locations of the wave potential's trough and crest, which is a very efficient mechanism generating chaos and violent mixing in the system with more than one particle.\cite{CaFi02}

For $P < 3/2$, the black line separates the cylinder into two domains: the orbits rotating (clockwise) around the elliptic fixed point, and the orbits winding (toward the left) around the cylinder. For the value $P=3/2$, the system has a saddle-center bifurcation at which an elliptic-hyperbolic pair coalesce, as shown by the black point in the bifurcation diagram Figs~\ref{fig:BifDiag} and \ref{fig:PxH}. The trajectory asymptotic to the bifurcation point is represented by the red line in Fig.~\ref{fig:03b}.

For $ P > 3 / 2$, we see two different types of stability at the same fixed point coordinate $y_{2,3}^*=\pi$. The upper fixed point is elliptic, circled by orbits rotating counterclockwise in the phase portrait. 
The lower fixed point is hyperbolic, and the lower branch of its separatrix winds (to the left) around the cylinder: from $t \to - \infty$, the particle leaves the crest of the wave potential, it passes at the bottom of the wave potential when the wave has its largest amplitude, and it asymptotes again the next crest of the potential for $t \to + \infty$, so that $\int_{-\infty}^{+\infty} \dot y(t) \rmd t = - 2 \pi$. 
The upper branch of the separatrix is, for $3/2 < P < 3 / 4^{1/3}$, a counterclockwise loop around the elliptic fixed point, with $\cos(x-\theta)$ always negative: from $t \to - \infty$, the particle leaves the crest of the wave potential, it passes again at the same crest of the wave potential when the wave has its smallest amplitude, and it asymptotes again the same crest of the potential for $t \to + \infty$, 
so that $\int_{-\infty}^{+\infty} \dot y(t) \rmd t = 0$. 

Fig.~\ref{fig:03c} shows another special value of $P$ for which the phase portrait changes: when $P = 3 / 4^{1/3}$, the points of null wave intensity belong to the separatrix of the X point ($y_2^* = \pi$, $p_2^*  \sqrt{2(P-p_2^*)} = 1$). The trajectory for which this happens has energy $H_{I=0} = P^2 / 2$ (for passing through $I = 0$), and this energy must also be equal to the energy of the X point, {\sl i.e.}  $H_{I=0} = (p_2^*)^2 /2 + \sqrt{2(P - p_2^*)}$ (where $p_2^*$ is the X point momentum obtained by solving condition $p_2^*  \sqrt{2(P-p_2^*)} = 1$). Thus, $P^2 = (p_2^*)^2 + 2 \sqrt{2(P - p_2^*)} = (p_2^*)^2 + 2 / p_2^*$, and the X point condition implies that $P = p_2^* + 1 / (2 (p_2^*)^2)$, so that $(p_2^*)^2  + 1 / p_2^* + 1 / (4 (p_2^*)^4) = (p_2^*)^2 + 2 / p_2^*$, {\sl i.e.} $p_2^* = 4^{-1/3}$ and $P = 3 / 4^{1/3}$. Because this phase portrait connects two distinct points (the $I = 0$ point and the X point), the system undergoes a global bifurcation at $P = 3 / 4^{1/3}$.

For $P > 3 / 4^{1/3}$, the upper branch of the separatrix winds around the cylinder (to the right): from $t \to - \infty$, the particle leaves the crest of the wave potential, it passes at the bottom of the wave potential when the wave has its smallest amplitude, and it asymptotes the next crest of the potential for $t \to + \infty$, so that $\int_{-\infty}^{+\infty} \dot y(t) \rmd t = 2 \pi$. Moreover, for $P > 3/4^{1/3}$, the black line containing the points with $I=0$ separates two domains: above it, trajectories circle counterclockwise around the elliptic fixed point at $y_3^* = \pi$, whereas trajectories wind around the cylinder (to the right) between it and the upper branch of the separatrix.

The system with $N=1$ particle coupled to $M=1$ wave does not generate chaos.  
In the next section, we describe the emergence of chaos by increasing the number of particles to $N=2$ in the single wave model. 
The symmetric case with $N=1$, $M=2$, although departing from the SWM, 
can be described by the same Hamiltonian (\ref{eqHXY}) and (\ref{eqHItheta}), 
and chaos occurs as soon as two waves with different phase velocities are present.\cite{ElEs03}

\section{The single wave and two particles}
\label{sec:M1N2}
\subsection{General aspects}
\label{sec:GeneralM1N2}
The $M=1$, $N=2$ system is the first step towards the dynamics of the paradigmatic single wave model, where the case of many particles sheds much light on fundamental plasma instabilities, in particular the bump-on-tail. The reference Hamiltonian $H_{\rm sc}^{2,1}$ (from now on, denoted simply $H$) 
\begin{subequations}
\begin{eqnarray}
  H 
  & = & \frac{p_1^2}{2} + \frac{p_2^2}{2} + \omega_0 \frac{X^2 + Y^2}{2} 
  \nonumber \\ &&
         + \varepsilon Y (\sin x_1 + \sin x_2) - \varepsilon X (\cos x_1 + \cos x_2),
   \label{eqH12XY} 
   \\
   & = & \frac{p_1^2}{2} + \frac{p_2^2}{2} + \omega_0 I
            \nonumber \\ && 
            - \varepsilon \sqrt{2 I} \, ( \cos (x_1 - \theta)  + \cos (x_2 - \theta) ),
  \label{eqH12Itheta}
\end{eqnarray}
\end{subequations}
describes two particles interacting self-consistently with one wave. Again, a Galileo transformation enables us to set $\omega_0 = 0$, leaving
\begin{subequations}
\begin{eqnarray}
  H 
  & = & \frac{p_1^2}{2} + \frac{p_2^2}{2}
        +  \varepsilon (Y (\sin x_1 + \sin x_2) + 
        \nonumber \\ && \qquad \qquad \quad
        - X (\cos x_1 + \cos x_2)),
   \label{eqH12XY0}
        \\
   & = & \frac {p_1^2} {2} + \frac {p_2^2} {2}
        - \varepsilon \sqrt{2 I} \, (\cos (x_1 - \theta) + \cos (x_2 - \theta) ).
\label{eqH12Itheta0}
\end{eqnarray}
\end{subequations}

Finally, rescaling all variables as $t' = \lambda^{-1} t$, $x' = x$, $\theta' = \theta$, $p' = \lambda p$, 
$I' = \lambda I$, $P' = \lambda P$, $X' = \lambda^{1/2} X$, $Y' = \lambda^{1/2} Y$, $H' = \lambda^2 H$ 
shows that the coupling parameter can also be scaled away with $\varepsilon' = \lambda^{3/2} \varepsilon$. 
Thus we are left with three cases: 
\begin{enumerate}
\item
$\varepsilon = 0$: the system is uncoupled;
\item
$\varepsilon = 1$: the coupling has unit strength and favors $x_{1, 2} \sim \theta$ energetically;
\item
$\varepsilon = -1$: the coupling has unit strength and favors $x_{1, 2} \sim \pi + \theta$ energetically, but this can be absorbed in the change of variable $\theta' = \theta + \pi$.
\end{enumerate}
The model is thus completely parametrized by total energy $H$ and total momentum $P = p_1 + p_2 + (X^2 + Y^2)/2$ for $\varepsilon = 1$. 
From here on, we set $\varepsilon = 1$. A similar Hamiltonian was considered by del Castillo Negrete and Firpo, \cite{Ca02,CaFi02} with a different wave-particle coupling. Our results complement theirs. 

For the $M = N = 1$ model, the fact that the wave intensity must be positive implied that the particle momentum $p$ was bounded from above by $P$. With two particles, total momentum $P$ sets no bound on a single particle momentum since only $p_1 + p_2$ is bounded by $P$. 

The original dynamics (\ref{eqH12XY}) or (\ref{eqH12XY0}) has three degrees of freedom, with phase space $(\TT \times \RR)^2 \times \RR^2$, where particles evolve on the cylinder $\TT \times \RR$ and the harmonic oscillator (\textit{viz}.\ the wave) evolves in the plane $\RR^2$. Given the two conserved quantities, the dynamics is restricted to 4-dimensional manifolds, and the motions generate Poincar\'e maps in 3-dimensional sections. 

The equations of motion read
\begin{subequations}
\begin{eqnarray}
  \dot x_r & = & p_r,
  \label{eq:dotx}
  \\
  \dot p_r & = & - X \sin x_r - Y \cos x_r = - \sqrt{2 I} \sin (x_r - \theta),
  \label{eq:dotp}
  \\
  \dot X & = & \sin x_1 + \sin x_2,
  \label{eq:dotX}
  \\
  \dot Y & = & \cos x_1 + \cos x_2,
  \label{eq:dotY}
  \\
  \dot \theta & = & - (2 I)^{-1/2} (\cos (x_1 - \theta) + \cos (x_2 - \theta)),
  \label{eq:dottheta}
  \\
  \dot I & = & \sqrt{2 I} (\sin (x_1 - \theta) + \sin (x_2 - \theta)).
  \label{eq:dotI}
\end{eqnarray}
\end{subequations}

For $\omega_0 > 0$, Hamiltonian (\ref{eqH12Itheta}) is bounded from below: $ | \cos(x_1 - \theta) + \cos(x_2 - \theta) | \leq 2$, so that
\begin{eqnarray}  
  H 
  & \geq & \frac{p_1^2 + p_2^2}2 - 2 \sqrt{2 I} + \omega_0 I,  \nonumber \\
  & = & \frac{p_1^2 + p_2^2}2 + \frac{\omega_0}2 (\sqrt{2 I} - \frac{2}{\omega_0})^2 - \frac{2}{\omega_0}.
  \label{eq:Hlowbound1}
\end{eqnarray}

For $\omega_0 \leq 0$, Hamiltonian (\ref{eqH12Itheta}) is not bounded from below: one may have $x_1 = x_2 = \theta$, $p_1 = p_2 = 0$ and $I$ arbitrarily large. Then $H =  \omega_0 I - 2 \sqrt{2I} \to - \infty$ as $I \to \infty$. However, for fixed $P$, the Hamiltonian is bounded from below even for $\omega_0 \leq 0$:
\begin{eqnarray}
  H  
  & \geq &\frac{p_1^2 + p_2^2}2 - 2 \sqrt{2 I} + \omega_0 I,  \nonumber \\
  & = &\frac{(p_1 + p_2)^2 + (p_1 - p_2)^2}4 - 2\sqrt{2} \sqrt{P - (p_1 + p_2)}
  \nonumber \\ && + \omega_0 (P - (p_1 + p_2)), \nonumber  \\
  & \geq & \frac{(p_1 + p_2)^2}4 - 2\sqrt{2} \sqrt{P - (p_1 + p_2)}              
  \nonumber \\   && 
               + \omega_0 (P - (p_1 + p_2)),
  \label{eq:Hlowbound2}
\end{eqnarray}
and the last two terms cannot diverge faster than the first one if $P$ is bounded. 

Given $P$ and $H$, equation (\ref{eq:Hlowbound2}) implies that $p_1$ and $p_2$ are bounded, 
and equation (\ref{eq:Hlowbound1}) shows that $I$ is bounded too. 
Since $x_1$, $x_2$ and $\theta$ vary on the unit circle, 
the constant $(P, H)$ manifolds are compact. 

This discussion about boundedness shows how important the conservation of momentum is. Moreover, it stresses how the notion of energy depends on the observer's viewpoint: a mere Galileo transformation changes the model from $H$ bounded from below for any $P$ (with $\omega_0 > 0$) to $H$ bounded from below conditionally on a fixed $P$.

\subsection{Reduction to 2 degrees of freedom}
\label{sec:PoincareVarM1N2}
The intersection of energy and momentum surfaces is compact for every $(H, P)$, for any fixed $\omega_0$. Indeed, the generalized coordinates $(x_1, x_2, \theta)$ range over a 3-torus. The generalized momenta must satisfy the above inequalities implying that 
neither $p_1$ nor $p_2$ can diverge, and hence $I = P - p_1 - p_2$ cannot diverge either. 

For a fixed $P$, consider the reduced dynamics in terms of $(y_1, y_2, p_1, p_2)$, with $I = P - p_1 - p_2$ and $y_r = x_r - \theta$. Then 
\begin{subequations}\label{eq:doty12}
\begin{eqnarray}
  \dot y_r & = & p_r + \frac{\cos y_1 + \cos y_2}{\sqrt{2 (P - p_1 - p_2)}},
  \label{eq:doty12a}
  \\
  \dot p_r & = & - \sqrt{2 (P - p_1 - p_2)} \sin y_r,
  \label{eq:doty12b}
\end{eqnarray}
\end{subequations}
with the conserved Hamiltonian
\begin{equation}
  H 
  = \frac {p_1^2} {2} + \frac {p_2^2} {2}
        -  \sqrt{2 (P - p_1 - p_2)} \, (\cos y_1 + \cos y_2)  . 
\label{eqH12py}
\end{equation}

As the $N = 1$ case is recovered by setting $y_1 = y_2, p_1 = p_2$ and rescaling time, energy and coupling constant, 
let $\sigma = P/2$. With variables $z_1 = (y_1 + y_2) / 2$, $z_2 = (y_1 - y_2) / 2$, $w_1 = (p_1 + p_2) / 2$, $w_2 = (p_1 - p_2) / 2$,
the Poisson brackets are
\begin{eqnarray}
  [f, g] 
  & = & \partial_{p_1} f \ \partial_{y_1} g - \partial_{y_1} f \ \partial_{p_1} g 
            + \partial_{p_2} f \ \partial_{y_2} g - \partial_{y_2} f \ \partial_{p_2} g,
  \nonumber \\
  & = & \frac{1}{2} (\partial_{w_1} f \ \partial_{z_1} g - \partial_{z_1} f \ \partial_{w_1} g    
             \nonumber \\ && \qquad
            + \partial_{w_2} f \ \partial_{z_2} g - \partial_{z_2} f \ \partial_{w_2} g ),
\end{eqnarray}
so that Hamilton's canonical evolution equations read
\begin{eqnarray}
  \dot g  = [H, g] 
  &=& (\partial_{w_1} \frac{H}{2}) \ \partial_{z_1} g - (\partial_{z_1} \frac{H}{2}) \ \partial_{w_1} g 
          \nonumber \\ 
   & &  + (\partial_{w_2} \frac{H}{2}) \ \partial_{z_2} g - (\partial_{z_2} \frac{H}{2}) \ \partial_{w_2} g \, .
\end{eqnarray}
Specifically,
\begin{subequations}
\begin{eqnarray}
  \dot z_1 
  & = & w_1 + \frac{\cos (z_1 + z_2) + \cos (z_1 - z_2)}{\sqrt{4 (\sigma - w_1)}},
  \nonumber \\
  & = & w_1 + \frac{\cos z_1 \cos z_2}{\sqrt{\sigma - w_1}},
  \label{eq:dotz1}  
  \\
  \dot z_2,
  & = & w_2,
  \label{eq:dotz2}  
  \\
  \dot w_1 & = & - \sqrt{4 (\sigma - w_1)} \ \frac{\sin (z_1 + z_2) + \sin (z_1 - z_2)}{2},
  \nonumber \\
  & = &   - 2 \sqrt{\sigma - w_1} \, \sin z_1 \cos z_2, 
  \label{eq:dotw1}
  \\
  \dot w_2 & = & - \sqrt{4 (\sigma - w_1)} \ \frac{\sin (z_1 + z_2) - \sin (z_1 - z_2)}{2},
  \nonumber \\
  & = &   - 2 \sqrt{\sigma - w_1} \, \cos z_1 \sin z_2.
  \label{eq:dotw2}
\end{eqnarray}
\end{subequations}

The new variables $(w_1, w_2, z_1, z_2)$ are not canonically equivalent to the original ones (since the bracket undergoes a rescaling by $1/2$), but the quantity
\begin{eqnarray}
  E = H / 2
  & = &  \frac{w_1^2}{2} + \frac{w_2^2}{2} 
        -  \sqrt{\sigma - w_1} \, (\cos (z_1 + z_2) + \cos (z_1 - z_2)),
  \nonumber \\
  & = &  \frac{w_1^2}{2} + \frac{w_2^2}{2} 
        -  2 \sqrt{\sigma - w_1} \, \cos z_1  \cos z_2,
\label{eqE12wz}
\end{eqnarray}
plays the role of a Hamiltonian in these new variables as the action differential of the system may be written as
\begin{eqnarray}
  \rmd S 
  & = & \sum_r p_r \rmd x_r + I \rmd \theta - H \rmd t, \nonumber
  \\
  & = & \sum_r p_r \rmd y_r + P \rmd \theta - H \rmd t, \nonumber
  \\
  & = & 2 \left( \sum w_r \rmd z_r + \sigma \rmd \theta - E \rmd t \right).
\end{eqnarray}
Note that $E = H/2$ is also the energy per particle, like $\sigma = P/2$ is the momentum per particle.

Energy $E$ can be rewritten in the form 
\begin{eqnarray}
  E =E_1 (w_1, z_1) + E_2 (w_2, z_2, w_1, z_1),
\end{eqnarray}
with
\begin{eqnarray}
  &&E_1 =\frac{w_1^2}{2} -  2 \sqrt{\sigma - w_1} \, \cos z_1,
  \\
  &&E_2 =\frac{w_2^2}{2} + 2 \sqrt{\sigma - w_1} \, \cos z_1 (1 - \cos z_2). 
\end{eqnarray}
This form extracts for $(w_1, z_1)$ an effective Hamiltonian $E_1$ which is the $N=1$ model, 
up to rescaling the coupling coefficient with a factor $\sqrt{2}$. 
The second term $E_2$ is positive if $\cos z_1 > 0$, which corresponds to the case where the two particles 
are not ``too far'' from each other, and describes their relative motion as that of a pendulum 
with time-dependent parameters. 

The periodic boundary conditions $y_r \equiv y_r + 2 \pi \!\!\mod (2 \pi)$ imply that the configuration space is a torus. 
The covering of this torus with cells of the form $z_r \equiv z_r + 2 \pi \!\!\mod (A_r)$ for an appropriate $A_r$ 
is not consistent if one sets $A_1 = A_2 = \pi$ for both $z_1$ and $z_2$. 
For the sake of safety, we set $A_1 = A_2 = 2 \pi$, 
which implies that the new cells have an area equal to twice that of the original ones, 
and two points in the cell $(z_1, z_2)$ correspond to a single point in $(y_1, y_2)$.

We analyze the Poincar\'e sections at $z_2 \equiv 0  \!\!\mod (2 \pi)$. 
For a given $\sigma$, a point $(w_1, z_1)$ in this section may correspond to different energies $E$, depending on $w_2$. 
More precisely, when both particles have the same $(p, y) = (p_1, y_1) = (p_2, y_2)$, we have the $N=1$ dynamics, with just a doubled mass and doubled coupling constant. This generates a family of solutions to the $N=2$ case. 
But for $N=2$ with $z_2 = 0$ and an arbitrary $w_2$, 
the two-particle case always has more energy than the $N=1$ case. 
Since $E$ is conserved, the initial excess energy $E_2 = w_2^2/2$ in the two-particle system 
may be taken as a perturbation parameter enabling chaos near the orbits of the integrable system. 

The $(z_2 = 0, w_2 = 0)$ trajectory appears in the Poincar\'e section $z_2 = 0$ as the boundary of the domain accessible for a given total energy $H$. Its stability is governed by the linearized equations
\begin{subequations}
\begin{eqnarray}
  \dot z_1 
  & = & w_1 + \frac{\cos z_1}{\sqrt{\sigma - w_1}},
  \label{eq:dotz1lin}  
  \\
  \dot w_1 
  & = &   - 2 \sqrt{\sigma - w_1} \, \sin z_1,
  \label{eq:dotw1lin}
  \\
  \delta \dot z_2 
  & = & \delta w_2,
  \label{eq:dotz2lin}  
  \\
  \delta \dot w_2 
  & = &   - ( 2 \sqrt{\sigma - w_1} \, \cos z_1 ) \delta z_2,
  \label{eq:dotw2lin}
\end{eqnarray}
\end{subequations}
where the $(w_1, z_1)$ dynamics is master and the $(w_2, z_2)$ dynamics is slave. Indeed, the Taylor expansion $\cos \delta z_2 = 1 - (\delta z_2)^2 / 2 + \ldots$ implies that $\delta z_2$ cannot appear in the $(w_1, z_1)$ dynamics. 
This master-slave description is, for $(\delta w_2, \delta z_2)$, a linearized version of Boozer's analysis of the emergence of chaos in Hamiltonian systems. \cite{Boozer1994}

If the $(w_1, z_1)$ trajectory remains confined in the band $\cos z_1 > 0$ (or $- \pi/2 < z_1 < \pi/2$), then the small perturbation $(\delta z_2, \delta w_2)$ obeys a linear evolution equation with time-periodic coefficients of Hill type, $\delta \ddot z_2 = - g(t) \delta z_2$ with a positive function $g(t)$. 
Though there may be resonances for some such $(w_1, z_1)$ trajectories, the perturbation may remain bounded. Indeed, the Poincar\'e sections show nice KAM tori in this range (Fig.~\ref{fig:04}), and one checks that $E_2$ is positive definite for $\cos z_1 > 0$. 

In contrast, when the $(w_1, z_1)$ trajectory enters the band $\cos z_1 < 0$ (or $- \pi/2 < z_1 - \pi < \pi/2$), 
then the perturbation obeys $\delta \ddot z_2 = - g(t) \delta z_2$ with a negative function $g(t)$. 
During this time, the perturbation is amplified (and the more as $z_1$ approaches $\pi$), 
and the system may leave the linear regime. 
Then $\cos z_2$ is no longer close to 1, and the relative motion $(w_2, z_2)$ feeds back upon the ``master'' variables. 
Such a process easily generates chaos, and one may expect that, soon enough, 
the trajectory approaches $z_1 \approx \pi$ and the associated hyperbolic point. 
As a result, one may expect a well-developed chaotic behavior for trajectories entering the band $\cos z_1 < 0$. 
Energetically, $E_2$ has an indefinite signature for $\cos z_1 < 0$.

The $(w_1, z_1)$ trajectories which come close to $w_1 = \sigma$ are also likely to behave chaotically, because this line corresponds to $I = 0$, and on this line the angle $z_1$ spontaneously jumps by $\pi$ to account for the sign reversal of both $X$ and $Y$ when the wave crosses null-amplitude. Then the corner of the wave cat's eye in original variables $(x_r, p_r)$ suddenly becomes its center, and conversely, which is a very efficient mixing process.\cite{Menyuk, CaFi02} 

\section{Fixed points and special trajectories}
\label{sec:FixPt_Traj_N2}

The equilibrium solutions for the case with $N = 2$ are given in terms of $(x_1, x_2, p_1, p_2)$ such that $(I, \theta)$ remain constant. Thus, if $I>0$ the amplitude and phase remain constant only if $x_2 = \pi + x_1 \, {\mathrm{mod}}\,(2 \pi)$, which implies $p_1 = p_2$. The latter requires $\sin (x_1 - \theta) = \sin (x_2 - \theta)$, which must therefore vanish, so that $x_1 = \theta \,{\mathrm{mod}}\, \pi$ and $p_1 = p_2 = 0$.  Then $P = I$ and $H = 0$. One particle stands on the unstable fixed point of the wave ($\pi + \theta$), while the other particle is at the bottom of the potential well ($\theta$). It has actually been proved that, for a finite-amplitude wave with fixed $(I, \theta)$, particles cannot move with respect to the wave, whatever their number $N$. \cite{Els01}

For $I = 0$, i.e.\  $X = Y = 0$, the amplitude remains $0$ only if $x_1 = \pi + x_2 \, {\mathrm{mod}}\,(2 \pi)$, implying $p_1 = p_2$ (which must not vanish), and then $P = 2 p_1$ and $H = P^2 / 4$. Both particles move at the same velocity and form a (two-particle) ballistic beam. Such solutions exist for any number $N \geq 2$ particles. \cite{EZE96} 

The limit case $I = 0$ (for which $\theta$ is undefined) in the first type 
is also the special case $P = 0$ in the second type. 
Moreover, if we seek regular solutions with merely a constant phase, say $\theta = 0$, with no loss of generality,  
this imposes $Y = 0$ and $\dot Y = 0$, hence $x_2 = \pi \pm x_1$, which has two solutions:
\begin{enumerate}
\item
  $x_2 = \pi + x_1$ implies $\dot X = 0$, which was considered hereabove: $X$ itself must be $0$.
\item
  $x_2 = \pi - x_1$ implies $p_2 = - p_1$, but it also implies $\sin x_1 = \sin x_2$, so that $\dot X = 2 \sin x_1$. However, total momentum $P = X^2/2$ must remain constant, which implies $\sin x_1 = 0$, hence $x_1 = 0$ and $x_2 = \pi$ (or the opposite).
\end{enumerate}
Thus, any other solution must have a time-dependent phase.

\subsection{Vanishing wave}
\label{sec:ZeroWaveN2}

For a vanishing wave amplitude, the special solution has 
its two particles moving at a finite velocity $v = p_1 = p_2 = P/2$.
Let then $x_1 = v t$ and $x_2 = \pi + v t$ (by a proper choice of the origin of time if $v \neq 0$). 
The resulting linearized dynamics reads 
{\scriptsize
\begin{equation}
   \frac{\rmd}{\rmd t}
   \left(\begin{matrix}
        \delta x_1 \cr  \delta p_1  \cr  \delta x_2  \cr  \delta p_2  \cr  \delta X  \cr  \delta Y
   \end{matrix}\right)
   = 
   \left(\begin{array}{c c c c c c}
        0         &      1      &         0          &      0      &          0          &         0            \cr  
        0         &      0      &         0          &      0      &     - \sin vt     &     - \cos vt      \cr  
        0         &      0      &         0          &      1      &          0          &         0            \cr  
        0         &      0      &         0          &      0      &       \sin vt     &       \cos vt      \cr  
     \cos vt   &      0      &     - \cos vt    &      0      &          0          &         0            \cr  
    - \sin vt   &      0      &       \sin vt     &      0      &          0          &         0       
   \end{array}\right)
   \left(\begin{matrix}
        \delta x_1 \cr  \delta p_1  \cr  \delta x_2  \cr  \delta p_2  \cr  \delta X  \cr  \delta Y
   \end{matrix}\right).
\end{equation}}
The explicit dependence of the matrix elements on time makes this linear dynamics a Floquet system. \cite{EZE96} 

One easily eliminates one pair of variables by introducing 
\begin{subequations}
\begin{eqnarray}
  s_1 & = & (\delta x_1 + \delta x_2) / 2,
  \\
  u_1 & = & \delta p_1 + \delta p_2,
  \\
  s_2 & = & \delta x_1 - \delta x_2, 
  \\
  u_2 & = & (\delta p_1 - \delta p_2) / 2,
\label{eq:newvar1}
\end{eqnarray}
\end{subequations}
so that the system decouples to 
{\small
\begin{equation}
   \frac{\rmd}{\rmd t}
   \left(\begin{matrix}
        s_1  \cr  u_1
   \end{matrix}\right)
   = 
   \left(\begin{array}{c c}
        0          &     1/2             \cr  
        0          &      0               \cr  
  \end{array}\right)
   \left(\begin{matrix}
        s_1  \cr  u_1
   \end{matrix}\right),
\end{equation}}
along with
{\small
\begin{equation}
   \frac{\rmd}{\rmd t}
   \left(\begin{matrix}
        s_2 \cr  u_2 \cr  \delta X  \cr  \delta Y
   \end{matrix}\right)
   = 
   \left(\begin{array}{c c c c c c}
        0         &      2      &          0          &         0            \cr  
        0         &      0      &     - \sin vt       &     - \cos vt        \cr  
     \cos vt      &      0      &           0         &         0            \cr  
    - \sin vt     &      0      &           0         &         0
   \end{array}\right)
   \left(\begin{matrix}
        s_2 \cr  u_2  \cr  \delta X  \cr  \delta Y
   \end{matrix}\right)   .
   \label{eq:dyn2}
\end{equation}}

Total momentum reads, to first order, $\delta P = u_1$, and energy to second order
\begin{equation}
   \delta^2 H = \frac{u_1^2}{4} + u_2^2 + (\delta X) s_2 \sin vt + (\delta Y) s_2 \cos vt   , 
\end{equation}
and the decoupling ensures that $u_1$ remains constant. Note that the perturbative approach does not require momentum conservation to second order, as the dynamics is linearized. The second order energy is relevant because the Hamiltonian is derived with respect to the perturbations to generate the dynamics. 

Though the decoupling reduces the number of dynamical variables in (\ref{eq:dyn2}), conservation laws no longer simplify the dynamics: total momentum conservation places no constraint on this system, and total energy is now formally time-dependent. 
However, since the coefficients in (\ref{eq:dyn2}) are trigonometric functions of time, one may view this system like forcing the wave oscillator by the particles reference motion, at  the Doppler-shifted angular frequency $v$. 

It is thus interesting to perform a Galileo transformation to the beam frame, introducing variables $x'_r = x_r - v t$, $p'_r = p_r - v$, $X' + \rmi Y' = (X + \rmi Y) \rme^{\rmi v t}$, so that (\ref{eq:dotx})-(\ref{eq:dotp})-(\ref{eq:dotX})-(\ref{eq:dotY}) read
\begin{subequations}
\begin{eqnarray}
  \dot x'_r & = & p'_r    ,
  \\
  \dot p'_r & = & - X' \sin x'_r - Y' \cos x'_r    ,
  \\
  \dot X' & = & \sin x'_1 + \sin x'_2 - v Y'    ,
  \\
  \dot Y' & = & \cos x'_1 + \cos x'_2 + v X'     .
\end{eqnarray}
\end{subequations}
The null solution now reads $x'_1 = 0$, $x'_2 = \pi$, $p'_1 = p'_2 = 0$, $X' = Y' = 0$, and the linearized dynamics is now autonomous,
{\small
\begin{equation}
   \frac{\rmd}{\rmd t}
   \left(\begin{matrix}
        \delta x'_1 \cr  \delta p'_1  \cr  \delta x'_2  \cr  \delta p'_2  \cr  \delta X'  \cr  \delta Y'
   \end{matrix}\right)
   = 
   \left(\begin{array}{c c c c c c}
        0         &      1      &         0          &      0      &          0          &         0            \cr  
        0         &      0      &         0          &      0      &          0          &        - 1          \cr  
        0         &      0      &         0          &      1      &          0          &         0            \cr  
        0         &      0      &         0          &      0      &          0          &         1            \cr  
        1         &      0      &        - 1         &      0      &          0          &        - v           \cr  
        0         &      0      &          0         &      0      &          v          &         0       
   \end{array}\right)
   \left(\begin{matrix}
        \delta x'_1 \cr  \delta p'_1  \cr  \delta x'_2  \cr  \delta p'_2  \cr  \delta X'  \cr  \delta Y'
   \end{matrix}\right)   .
   \label{eq:dyn4}
\end{equation}}
Again, introducing $s_1, s_2, u_1, u_2$ decouples the system, and the nontrivial part reads
{\small
\begin{equation}
   \frac{\rmd}{\rmd t}
   \left(\begin{matrix}
        s_2 \cr  u_2 \cr  \delta X'  \cr  \delta Y'
   \end{matrix}\right)
   = 
   \left(\begin{array}{c c c c c c}
        0         &      2      &          0          &         0            \cr  
        0         &      0      &          0          &        - 1           \cr  
        1         &      0      &          0          &        - v           \cr  
        0         &      0      &          v          &         0       
   \end{array}\right)
   \left(\begin{matrix}
        s_2 \cr  u_2  \cr  \delta X'  \cr  \delta Y'
   \end{matrix}\right)   .
   \label{eq:dyn5}
\end{equation}}

The characteristic polynomial of the matrix in the right hand side of (\ref{eq:dyn4}) is 
\begin{equation}
  P_4(\lambda) = \lambda^4 + v^2 \lambda^2 + 2 v     ,
  \label{eq:pol1}
\end{equation}
which admits four roots,
\begin{equation}
  \lambda = \pm \frac{\rmi}{\sqrt{2}} \sqrt{ v^2 \pm \sqrt{v^4 - 8 v}}     .    
  \label{eq:pol2}
\end{equation} 
If $v < 0$, viz.\ if the particles are slower than the wave, this defines two real roots and two imaginary roots: the reference solution is unstable. 
If $0 < v < 2$, viz.\ if the particles are moderately faster than the wave, this defines a quartet of complex roots
$\pm \alpha \pm \rmi \beta$:
the reference solution is also unstable. 
If $v > 2$, viz.\ if the particles are significantly faster than the wave, all four roots are purely imaginary: 
the reference state is stable. 

At $v = 0$, the case coincides with the zero-amplitude limit of the other class of solutions. 
For $v$ close to 0 and negative, the eigenvalues have modulus $(- 2 v)^{1/4}$ and lie on the real and imaginary axes.
For $v$ close to 0 and positive, the eigenvalues have modulus $(2 v)^{1/4}$ and lie on the bissectrices. 
At $v = 2$, the model exhibits a Kre{\u\i}n collision (or Hamiltonian Hopf bifurcation). \cite{Meiss}

\subsection{Coinciding particles}
\label{sec:ReducN2N1}

If the two particles are at the same position with the same velocity at $t=0$, 
they will never separate. Indeed, with variables $(p_r, x_r, Y, X)$, 
the evolution equations (\ref{eq:dotx})-(\ref{eq:dotp})-(\ref{eq:dotX})-(\ref{eq:dotY}) 
are smooth and have unique solutions globally in time. 

For $z_2 = 0, w_2 = 0$, the evolution equations (\ref{eq:dotz1lin})-(\ref{eq:dotw1lin}) with time $t$
can be rewritten as (\ref{eq:doty})-(\ref{eq:dotv}) with a time $s$ by rescaling 
$s = 2^{1/3} t$, $y = z_1$, $p = 2^{-1/3} w_1$, $P_{N=1} = 2^{-1/3} \sigma$ and $H_{N=1} = 2^{-2/3} H_{N=2}$. 
Note that the rescaling of time implies that frequencies and Lyapunov exponents rescale accordingly. 

The wave amplitude $I>0$ implies that $p_1 + p_2 < P$, then $\theta$ is well-defined, 
and we have $\sin y_1 = \sin y_2 = 0$. 
So when both particles are in the same position $y_1=y_2=0$, from (\ref{eq:doty12}) one finds 
\begin{eqnarray}
   p_1 = p_2 = \frac{-1}{\sqrt{\sigma - p_1}} < 0.
\end{eqnarray}
The linear stability analysis from eqs. (\ref{eq:doty12}), 
\begin{subequations}
\begin{eqnarray}
   \delta \dot{y}_1 & = & \delta p_1, \\
   \delta \dot{p}_1 & = & \frac{2}{p_1} \delta y_1, \\
   \delta \dot{y}_2 & = & \delta p_2, \\
   \delta \dot{p}_2 & = & \frac{2}{p_1} \delta y_2,
\end{eqnarray}
\end{subequations}
shows that this case is stable, so that both particles oscillate at the same frequency near the bottom of the wave potential well. 
As we see on Figs~\ref{fig:04}, \ref{fig:05} and \ref{fig:06}, 
at low energy the system undergoes harmonic oscillations near the fixed point at $y_1 = y_2 = 0$. 
For increasing energy, the nonlinear coupling generates chaos near this 1:1 resonance. 

Now considering the case where both particles coincide at the same position $y_1 = y_2 = \pi$ 
(unstable position of the potential), 
\begin{equation}
    p_1 = p_2 = \frac{1}{\sqrt{\sigma - p_1}} > 0,
\end{equation}
this solution exists only if $\sigma \ge 3/4^{1/3}$. 
Here we have two possibilities, namely 
$p_1 = p_2 = p^*_{\mathrm{low}}$ and $p_1 = p_2 = p^*_{\mathrm{high}}$.

The linearized equations read
\begin{subequations}
\begin{eqnarray}
   \delta \dot{p}_r & = & 2\sqrt{\sigma - p^*} \delta \dot{y}_r, \\
   \delta \dot{y}_r & = & \delta p_r - \frac{1}{4(\sigma - p^*)^{3/2}} (\delta p_1 - \delta p_2),
\end{eqnarray}
\end{subequations}
or 
{\small
\begin{equation}
   \left(\begin{matrix}
        \delta \dot{y}_1 \cr  \delta \dot{y}_2  \cr  \delta \dot{p}_1  \cr  \delta \dot{p}_2
   \end{matrix}\right)
   = 
   \left(\begin{array}{c c c c c c}
        0              &      0                &         1-K      &      -K      \cr  
        0              &      0                &         -K       &      1-K     \cr  
   2\sqrt{\sigma-p^*}  &      0                &          0       &       0      \cr  
        0              &   2\sqrt{\sigma-p^*}  &          0       &       0
        
   \end{array}\right)
   \left(\begin{matrix}
        \delta y_1 \cr  \delta y_2  \cr  \delta p_1  \cr  \delta p_2
   \end{matrix}\right),    
\end{equation}}
with
\begin{eqnarray}
   K = \frac{1}{4 (\sigma - p^*)^{3/2}} = p^{*3} / 4.
\end{eqnarray}
Thus
{\small
\begin{equation}
   \left(\begin{matrix}
        \delta \dot z_2 \cr  \delta \dot z_1  \cr  \delta \dot w_2  \cr  \delta \dot w_1 
   \end{matrix}\right)
   = 
   \left(\begin{array}{c c c c c c}
        0              &      0                &         1       &      0       \cr  
        0              &      0                &         0       &     1-2K     \cr  
   2\sqrt{\sigma-p^*}  &      0                &         0       &       0      \cr  
        0              &   2\sqrt{\sigma-p^*}  &         0       &       0        
   \end{array}\right)
   \left(\begin{matrix}
        \delta z_2 \cr  \delta z_1  \cr  \delta w_2  \cr  \delta w_1 
   \end{matrix}\right),
\end{equation}}
and the dynamics decouples $(\delta z_1, \delta w_1)$ and $(\delta z_2, \delta w_2)$. 

The ``center of mass'' dynamics $(\delta z_1, \delta w_1)$ near $y_1 = y_2 = \pi$ gives
\begin{subequations}
\begin{eqnarray}
  \delta \dot{z_1} & = & (1 - 2K) \delta w_1, \\
  \delta \dot{w_1} & = & 2\sqrt{\sigma -p^*} \delta z_1.
\end{eqnarray}
\end{subequations}
The eigenvalues are given by
\begin{eqnarray}
   \lambda^2 = 2\sqrt{\sigma - p^*}\bigg(1- \frac{1}{2} p^{*3}\bigg) . 
\end{eqnarray}
They are real for $p^* < 2^{1/3}$ and imaginary for $p^* > 2^{1/3}$. 
Thus the $p^*_{\mathrm{low}}$ solution is unstable and the $p^*_{\mathrm{high}}$ one is stable. 
The critical value $p^* = 2^{1/3}$ implies that $\sigma = 3/4^{1/3}$, in agreement with the $N=1$ case. 

For the ``relative motion'' $(\delta z_2, \delta w_2)$ dynamics 
\begin{eqnarray}
   \delta \dot{z}_2 & = & \delta w_2, \\
   \delta \dot{w}_2 & = & 2 \sqrt{\sigma - p^*} \delta z_2,
\end{eqnarray} 
the eigenvalues solve $\lambda^2 = 2\sqrt{\sigma -p^*} >0$ and are always real. 
Therefore the coinciding particles solution at $y_1 = y_2 = \pi$ with $p_1 = p_2$ is always unstable.

\subsection{Oppositely placed particles}
\label{sec:BallisticN2}

For the nonvanishing wave reference state, let $\theta_0 = 0$, 
viz.\ $X_0 = \sqrt{2 I_0}$ and $Y_0 = 0$, $x_{10} = 0$, $x_{2 0} = \pi$, $p_{10} = p_{20} = 0$. 
This exact solution has no analogue in the $N=1$ case. 
Near this state, the energy reduces to second order to 
\begin{equation}
  \delta^2 H 
  = \frac{\delta p_1^2 + \delta p_2^2}2 
     + \delta Y (\delta x_1 - \delta x_2) 
     + X_0 \frac{\delta x_1^2 - \delta x_2^2}2,
\label{eq:d2HX0}
\end{equation}
and momentum to first order to 
\begin{equation}
   \delta P = \delta p_1 + \delta p_2 + X_0 \delta X.
\label{eq:dPX0}
\end{equation}
The linearized evolution equations read 
{\small
\begin{equation}
   \frac{\rmd}{\rmd t}
   \left(\begin{matrix}
        \delta x_1 \cr  \delta p_1  \cr  \delta x_2  \cr  \delta p_2  \cr  \delta X  \cr  \delta Y
   \end{matrix}\right)
   = 
   \left(\begin{array}{c c c c c c}
        0     &      1      &      0      &      0      &      0      &      0       \cr  
     - X_0  &      0      &      0      &      0      &      0      &     - 1      \cr  
        0     &      0      &      0      &      1      &      0      &      0       \cr  
        0     &      0      &     X_0   &      0      &      0      &      1       \cr  
        1     &      0      &     - 1     &      0      &      0      &      0       \cr  
        0     &      0      &      0      &      0      &      0      &      0       
   \end{array}\right)
   \left(\begin{matrix}
        \delta x_1 \cr  \delta p_1  \cr  \delta x_2  \cr  \delta p_2  \cr  \delta X  \cr  \delta Y
   \end{matrix}\right)   . 
\end{equation}}
The eigenvalue 0 is degenerate, with eigenvector $(1, 0, 1, 0, 0, - X_0)^{\mathrm T}$ 
corresponding to a simple translation in space and the associated change in the wave phase, 
and eigenvector $(0, 0, 0, 0, 1, 0)^{\mathrm T}$ corresponding to a change in the wave intensity. 
The latter eigenvector changes $P$. Neither eigenvector changes $H$. 

The eigenvalues $\lambda = \pm \sqrt{X_0}$ are simple, 
with eigenvectors $(0, 0, \lambda, X_0, -1, 0)^{\mathrm T}$ 
corresponding to particle 2 moving in the vicinity of its unstable equilibrium (conditioned by the wave). 
The change of momentum for particle 2 is compensated with the change of wave intensity,
so that these two (complex conjugate) eigenvectors lie in the plane tangent to constant $(P, H)$ surfaces.  

The eigenvalues $\lambda = \pm \rmi \sqrt{X_0}$ are simple, 
with eigenvectors $(\lambda, - X_0, 0, 0, 1, 0)^{\mathrm T}$ corresponding 
to particle 1 oscillating in the vicinity of its stable equilibrium (conditioned by the wave). 
The change of momentum for particle 1 is compensated with the change of wave intensity, 
so that these two eigenvectors lie in the plane tangent to constant $(P, H)$ surfaces.   

In summary, the $I > 0$ solutions are unstable in 6-dimensional space: they have two eigenvectors with 0 eigenvalue (cf. constants of the motion), two eigenvectors related to elliptic perturbations, and two related to hyperbolic perturbations. 

The clear link between the four nonzero eigenvalues and the motion of a single particle 
should not obscure the fact that the wave variables $(X, Y)$ must also evolve during these eigenmotions. 
Indeed, the particle-wave system is self-consistent, 
and one should not use blindly the stability analysis relevant to \emph{slaved} particles 
(though this analysis hints at the actual self-consistent behaviour). 

Finally, let us recall that this analysis is formulated for the fixed point $y_1 = 0, y_2 = \pi, p_1 = p_2 = 0$. 
On relabelling particles, it also applies to the fixed point $y_1 = \pi, y_2 = 0, p_1 = p_2 = 0$. 
In the 4-dimensional phase space of the reduced model (\ref{eq:doty12}) 
with fixed total momentum $P$, 
this latter fixed point is distinct from the former one. 
Therefore, the stable and unstable manifolds of both fixed points will generate heteroclinic connections 
within their common homoclinic tangle.

\section{Regular and chaotic trajectories}
\label{sec:TrajectoriesM1N2}

Chaos in the self-consistent interaction of two particles $(N=2)$ with one wave $(M=1)$ is expected, 
since this is a non-integrable Hamiltonian system, and there is no nontrivial solution with a traveling wave.\cite{Els01} 
Moreover, it is intuitive to think that typically chaos starts 
and is more intense in the regions close to the separatrix of the $N=1$ system.\cite{ElEs03} 
In particular, the explicit solution for the separatrix can be used to prove nonintegrability 
of perturbations of this system using the Melnikov-Poincar\'e integral. \cite{kozlov1983,GuckHol83} 

In our case, there are two standard ways for chaos to appear and grow. 
One is the homoclinic tangle growing from a separatrix, and the other is resonances near elliptic points, 
as discussed in the previous section. 

To keep the discussion simple, we consider here the case $P = 2$, so that $\sigma = 1$, 
and the $N=1$ reference model has $P_{N=1} = 2^{-1/3}$. 
For this total momentum, the integrable $N = M = 1$ system has only one fixed point, 
as in Fig.~\ref{fig:03a}.

\subsection{Chaos near the elliptic fixed point for $H < 0$ }
\label{sec:Chaos_near_elliptic_point}

\begin{figure}[!tb]
  \centering
   \includegraphics[width=0.4\textwidth]{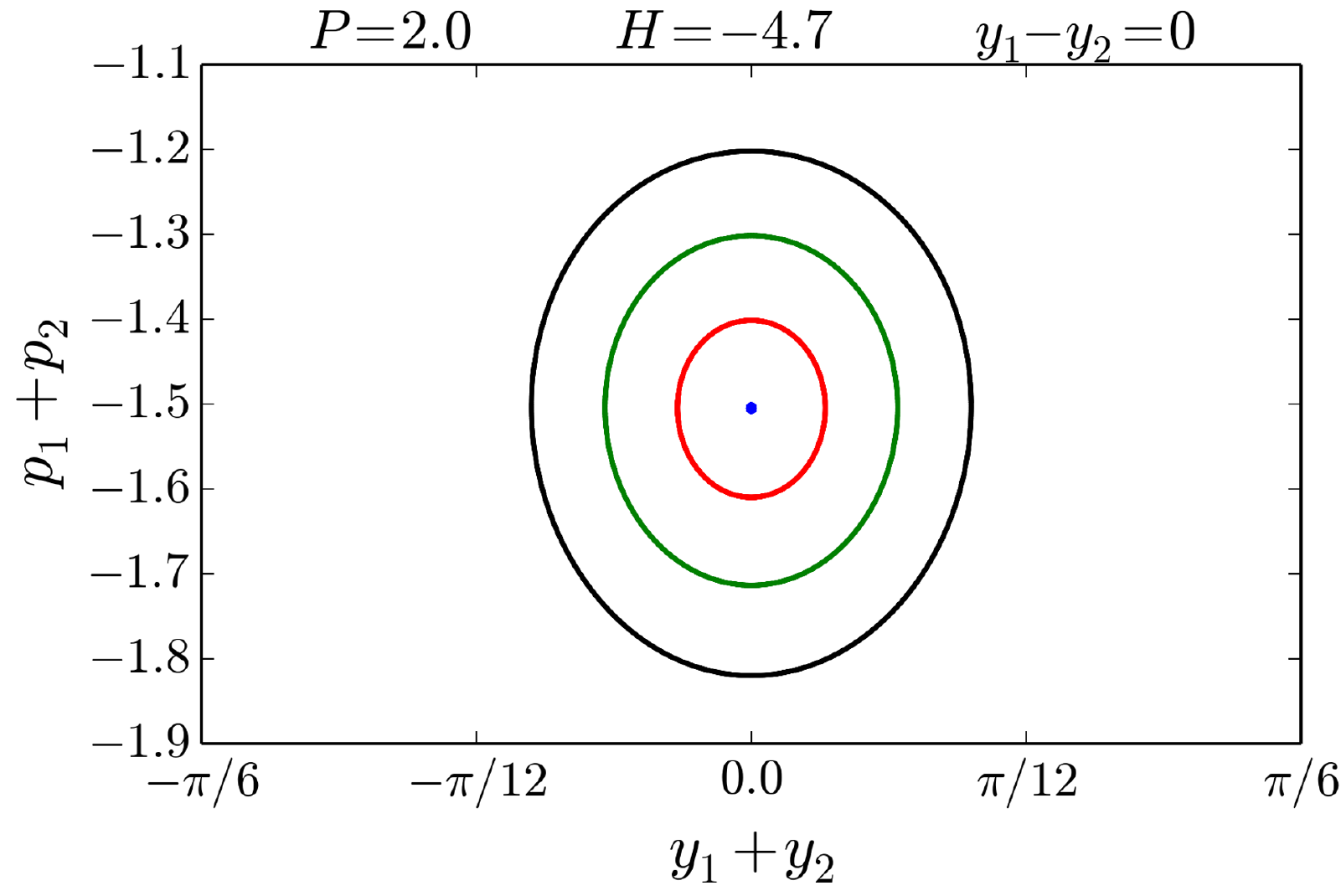}
   \caption{Interception of trajectories with the Poincar\'e section located at $y_1 - y_2 = 0$.  
   This panel represents the dynamics in the neighbourhood of the elliptic fixed point at $y_1 - y_2 = 0$ 
   with total energy $H = -4.7$.}
   \label{fig:04}
\end{figure}

In the negative energy regime, the wave intensity is large and the kinetic energy of the particles is low, 
so that the particles oscillate at the bottom of the wave potential well.
For a very small perturbation, the two particles ``agglomerate'' and move together 
in such a way that the evolution can be understood as if there was only one particle $(N = 1)$ in the system. 
This dynamics is represented by the black (outer oval) trajectory in Fig.~\ref{fig:04}. 

The perturbation strength, which is given by the difference in the initial velocities of the particles, 
increases from the black (outer oval) to the blue (central point) trajectory. 
The outer oval trajectory corresponds to $w_2 = 0$, $z_2 = 0$, and it remains forever in the Poincar\'e section plane, as we saw in Section~\ref{sec:PoincareVarM1N2}. 
The other trajectories only intersect the section plane at times at which the two particles cross each other (having then a nonzero relative velocity $2 w_2$). 

\begin{figure}[!tb]
  \centering
   \includegraphics[width=0.4\textwidth]{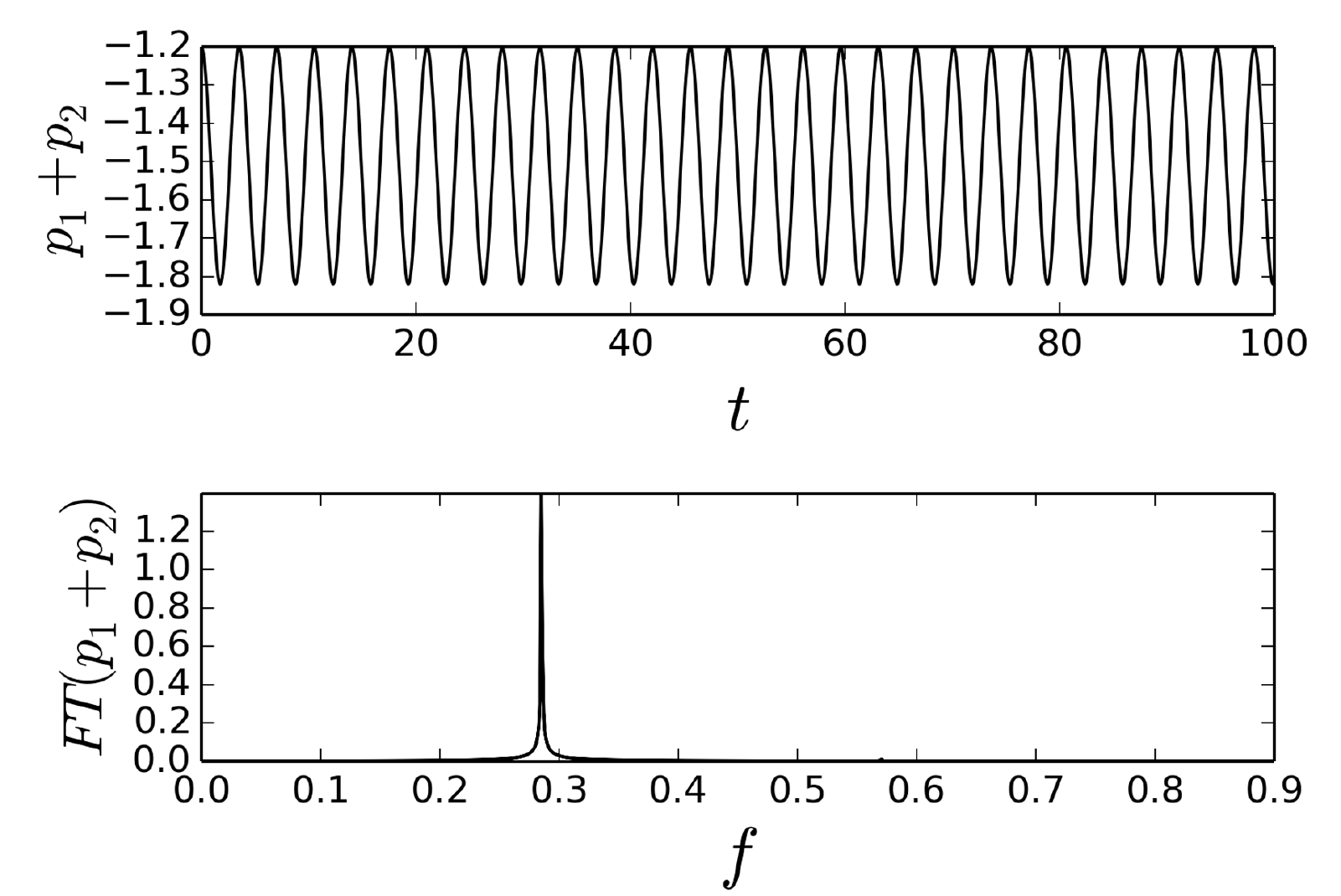}
   \caption{Time evolution and Fourier transform of the particles total momentum for the  black (outer oval) trajectory in Figure~\ref{fig:04}.}
   \label{fig:05}
\end{figure}

The Fourier transform of the particles total momentum $p_1 + p_2$ for the black (outer oval) trajectory is displayed in Fig.~\ref{fig:05}, 
and it shows that the system then oscillates harmonically with a single frequency.
 
\begin{figure}[!tb]
  \centering
   \includegraphics[width=0.4\textwidth]{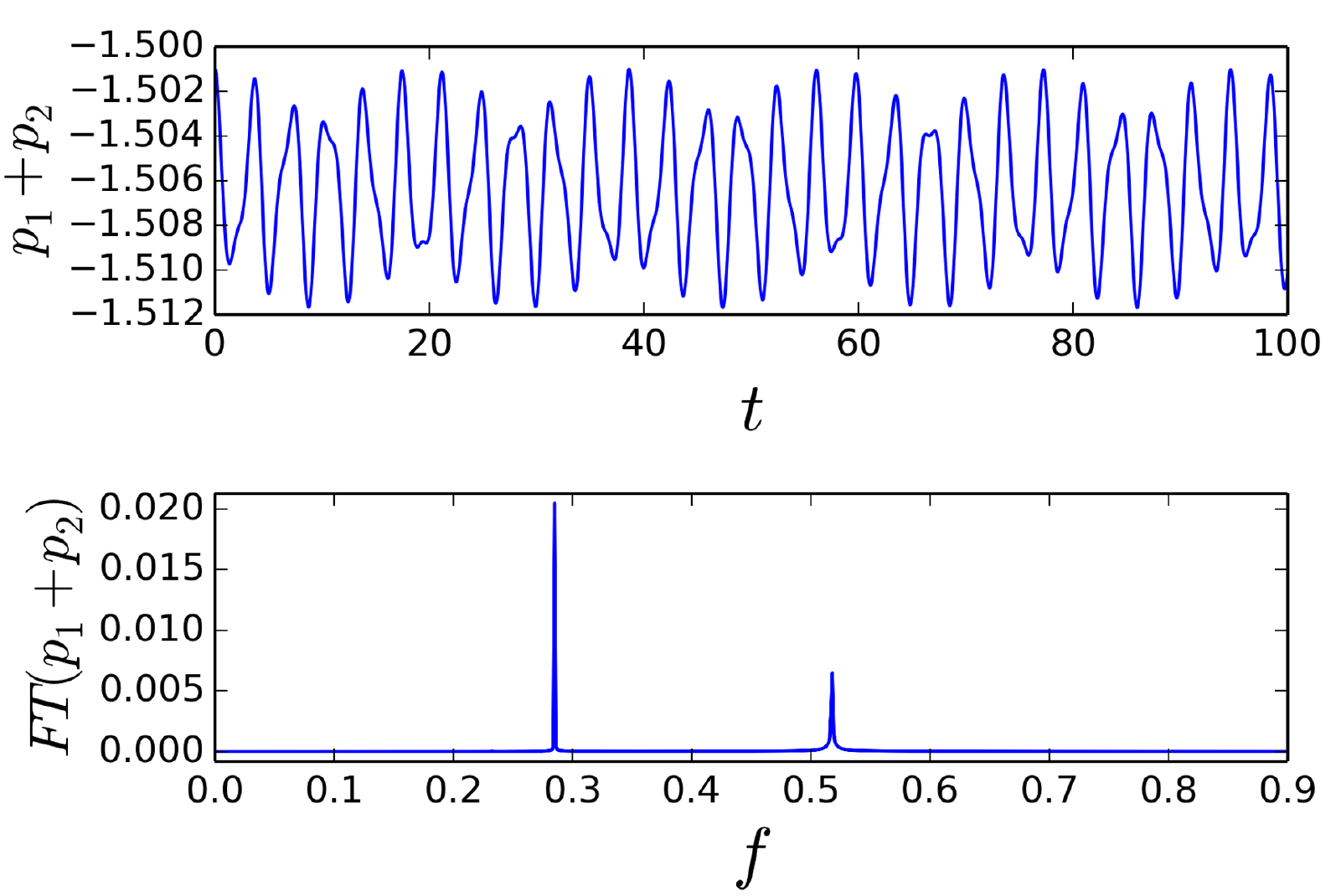}
   \caption{Time evolution and Fourier transform of the particles total momentum for the blue (central point) trajectory in Figure~\ref{fig:04}.}
   \label{fig:06}
\end{figure}

\begin{figure}[!tb]
  \centering
   \subfigure[]{\includegraphics[width=0.4\textwidth]{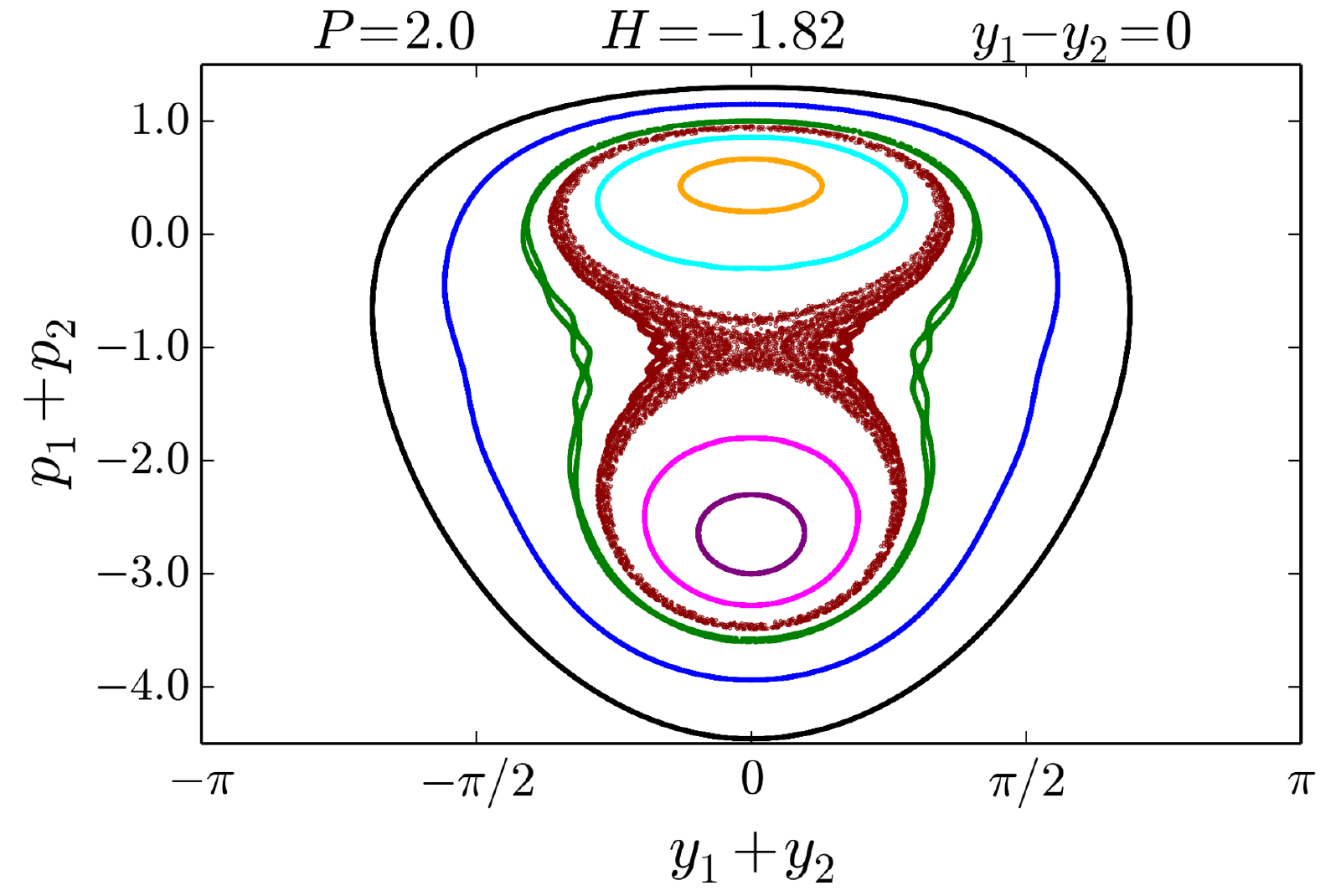}\label{fig:07a} }
   \subfigure[]{\includegraphics[width=0.4\textwidth]{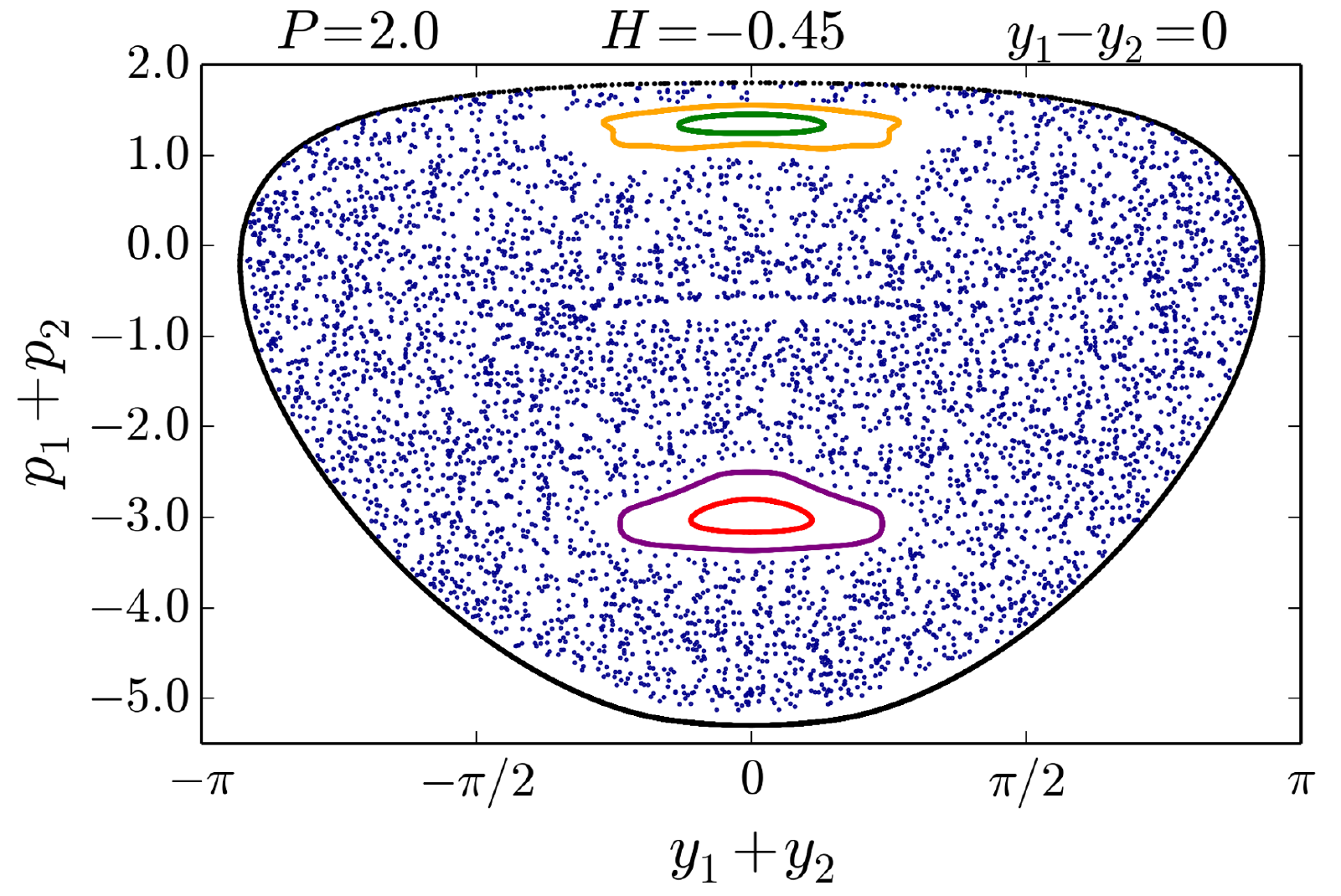}\label{fig:07b} }
   \caption{Interception of trajectories with the Poincar\'e section located at $y_1 - y_2 = 0$ for negative $H$ values.
    The panels represent the dynamics in the neighbourhood of the elliptic fixed point at $y_1 - y_2 = 0$ 
    with total energy (a) $H = -1.82$, and (b) $H = -0.45$.}
   \label{fig:07}
\end{figure}

For the blue (central point) trajectory in Figure~\ref{fig:04}, which has the highest perturbation strength for this energy surface, 
we find that, as we increase the disturbance in the system, 
the oscillation amplitude of the particles center of mass increases 
and the particles start oscillating in anti-phase with respect to each other. 
The relative motion of the particles with respect to the wave gives rise to a resonance, 
as shown by the Fourier transform in Figure~\ref{fig:06}. 
The contribution of this resonance is eventually enough to establish resonance overlap and chaos, 
as seen in Figures~\ref{fig:07a} and \ref{fig:07b}, with ($w_1, z_1$) trajectories confined in the band $\cos z_1 > 0$.

This is the usual scenario near an elliptic fixed point, 
with deformation and destruction of tori due to increased disturbance as predicted by the KAM theorem.\cite{Ott2002} 
Furthermore, the Poincar\'e-Birkhoff theorem predicts that 
when a resonant torus is destroyed (due to the increase in the perturbation), 
a sequence of periodic orbits 
will appear in phase space, 
which alternate between elliptic (stable) and hyperbolic (unstable),  
generating periodic points in the Poincar\'e section. 
In this scenario, hyperbolic points are related to the emergence of chaos, 
while elliptic points become the center of stable regions, called resonant islands, 
immersed in the chaotic sea. \cite{Ott2002} 
When the perturbation is increased, the trajectories that contain an unstable point 
(as the one similar to an "8" in Fig.~\ref{fig:07a}) 
give rise to chaos in this region. 

\begin{figure}[!tb]
  \centering
   \subfigure[]{\includegraphics[width=0.35\textwidth]{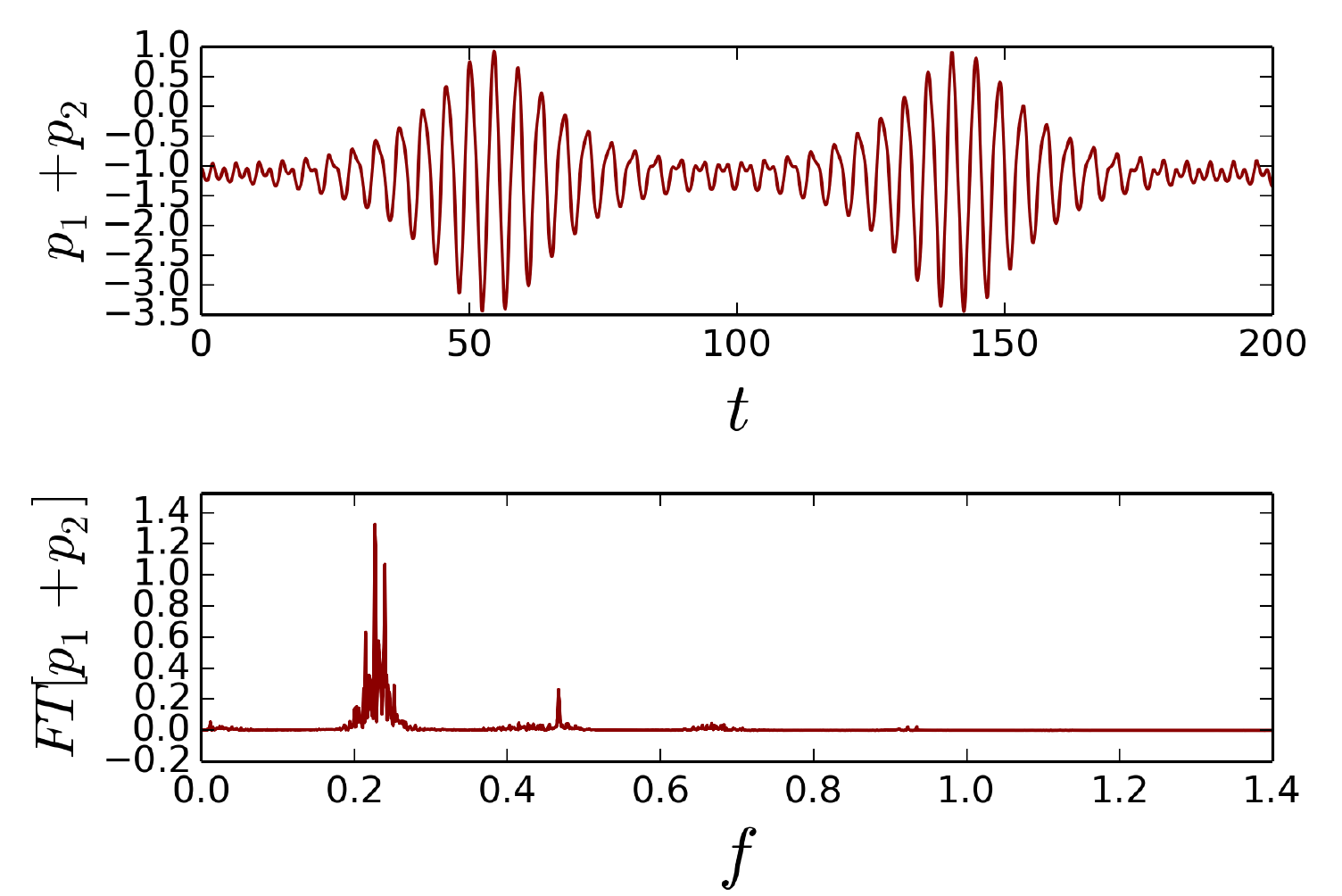}\label{fig:08a}} \\
   \subfigure[]{\includegraphics[width=0.35\textwidth]{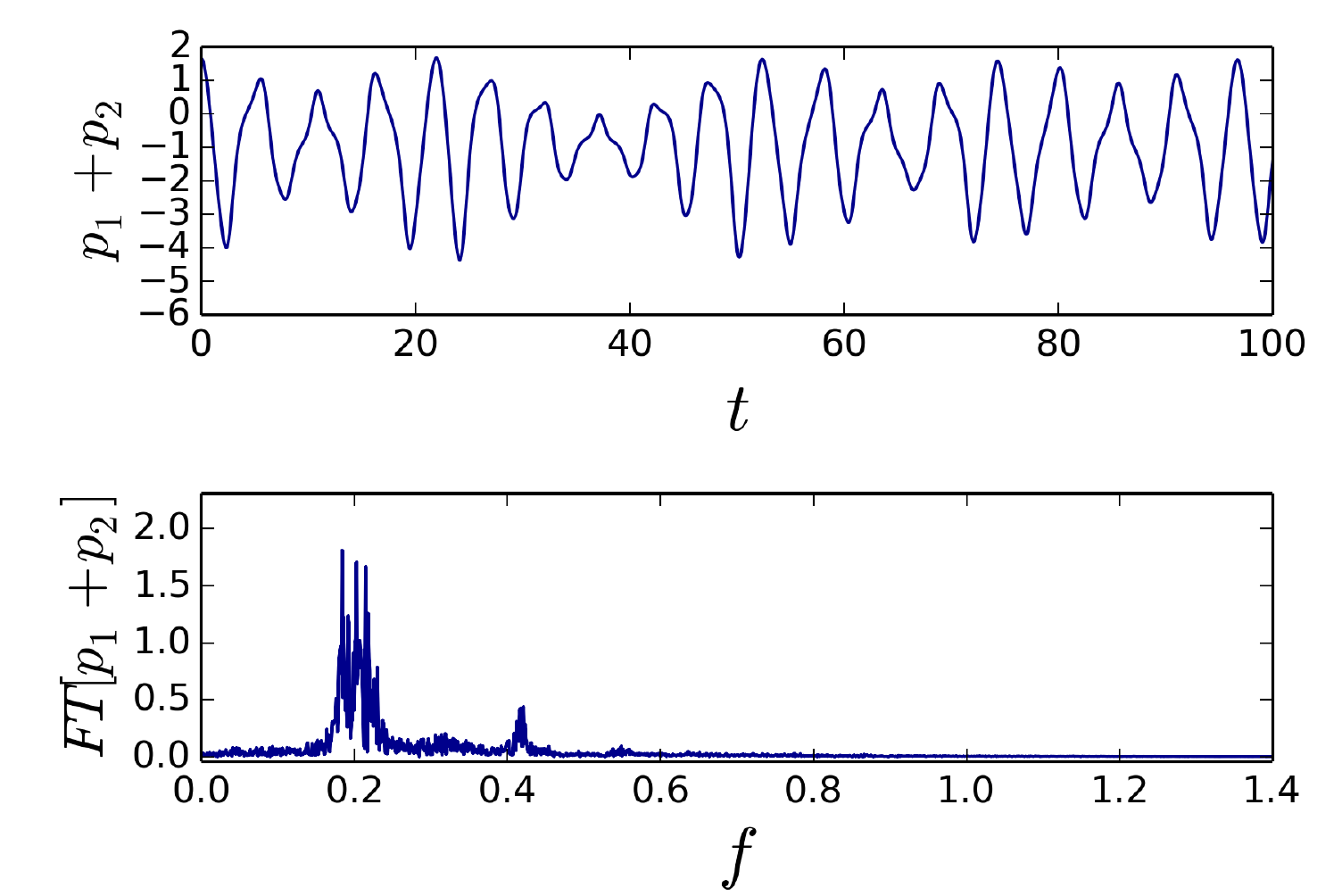}\label{fig:08b}}
   \caption{Time evolution and Fourier transform of the particles total momentum for (a) the dark red (similar to an "8") trajectory in Figure~\ref{fig:07a}, and (b) the blue (chaotic) trajectory in Figure~\ref{fig:07b}. }
   \label{fig:08}
\end{figure}

Figures~\ref{fig:08a} and \ref{fig:08b} show the time evolution of the particles total momentum 
and its Fourier transform for the dark red (similar to an "8") and the blue (chaotic) trajectories of Figs~\ref{fig:07a} and \ref{fig:07b}, respectively. 
Despite the noise in the Fourier-transformed signal, the peak frequency and its harmonics still appear well defined. 
This may be related to the fact that when one or both particles escape from the wave potential well 
(giving rise to a burst of chaos), they are easily recaptured by the wave potential well. 
Hence, in this energy regime, the chaotic trajectory does not present a large excursion through phase space. 

\subsection{Chaos with a separatrix for $0 \leq H < P^2 / 4$ }
\label{sec:Chaos_near_separatrix}

\begin{figure}[!tb]
  \centering
   \subfigure[]{\includegraphics[width=0.4\textwidth]{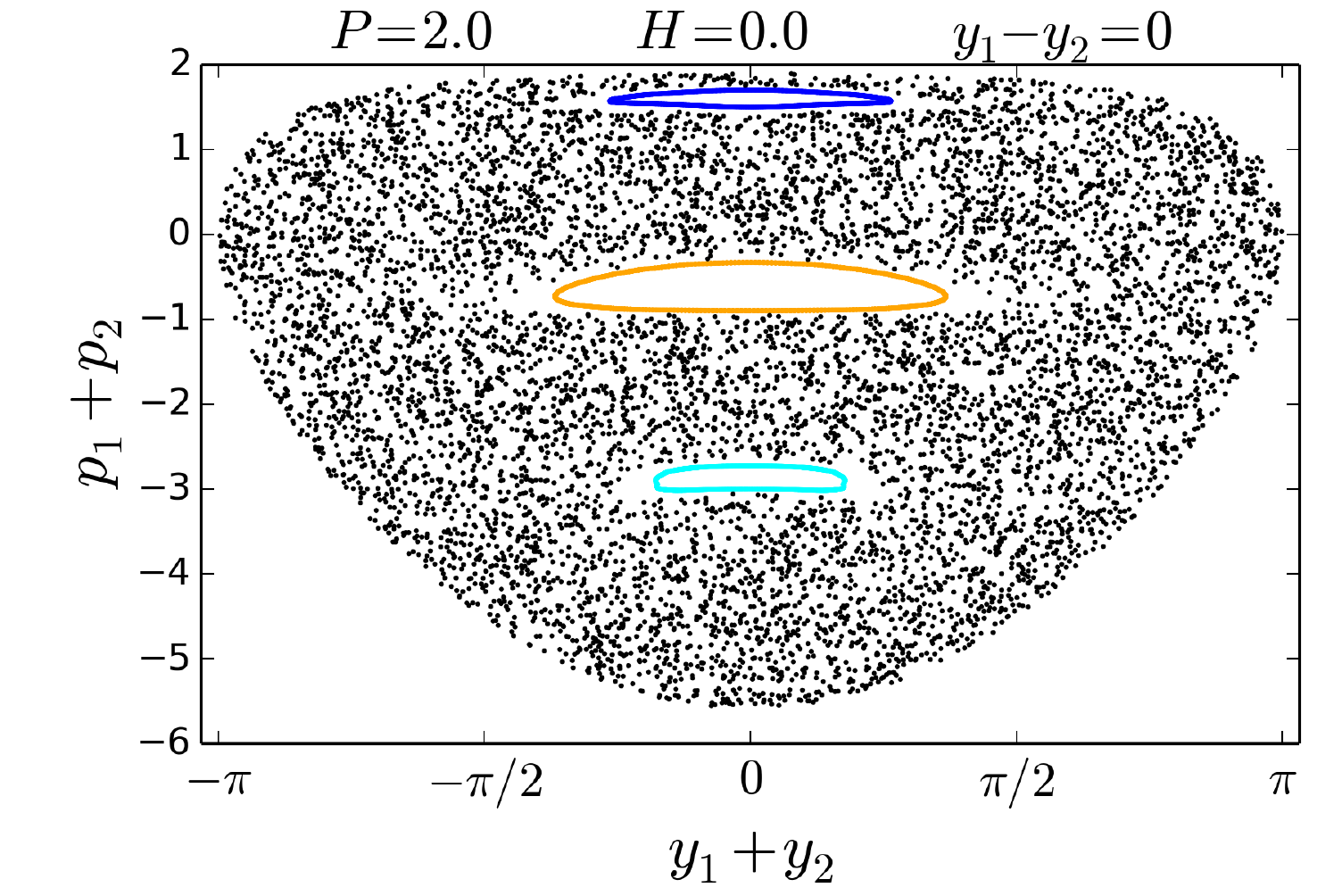}\label{fig:09a}}
   \subfigure[]{\includegraphics[width=0.4\textwidth]{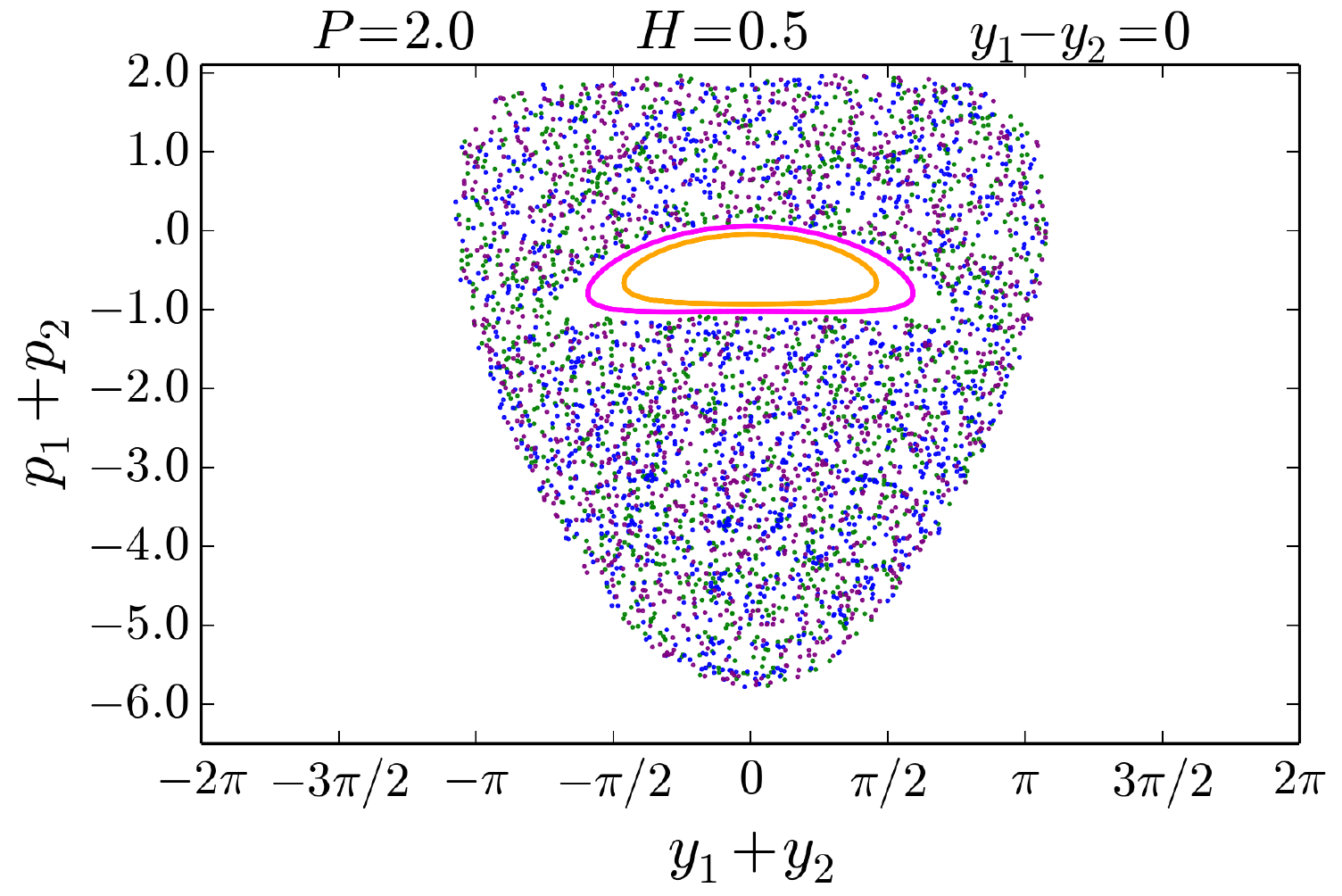}\label{fig:09b}}
   \subfigure[]{\includegraphics[width=0.4\textwidth]{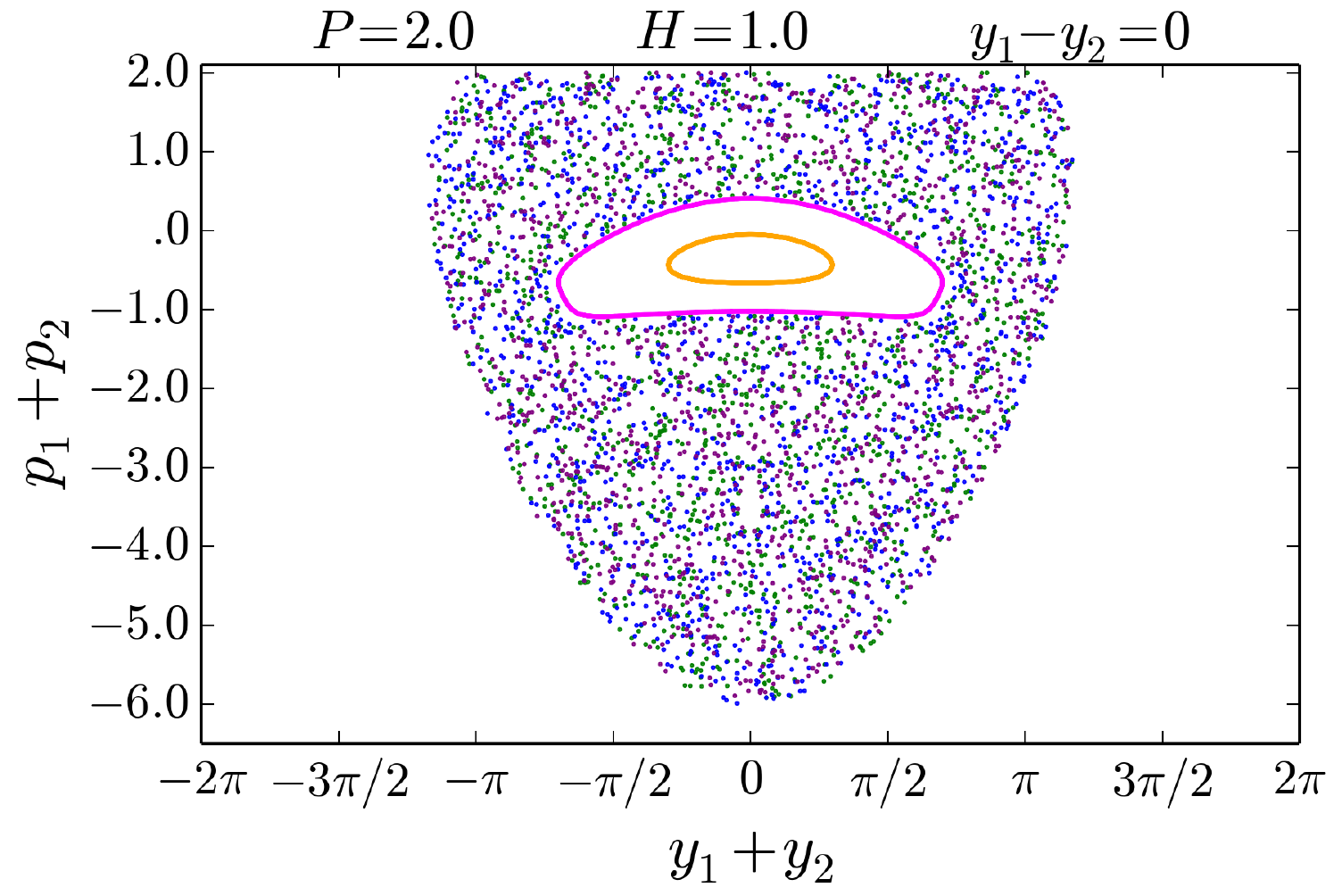}\label{fig:09c}}
   \caption{Poincar\'e section at $y_1 - y_2 = 0$ for moderate positive $H$ values. (a) $H=0.0$, (b) $H=0.5$, and (c) $H=1.0$.}
   \label{fig:09}
\end{figure}

At $H=0$, for positive $P$, a new special solution appears, 
with particles at opposite positions (see Sec.~\ref{sec:BallisticN2}). 
This solution is ``far'' from the Poincar\'e surface $y_1 = y_2$, 
and it does not significantly alter the Poincar\'e sections,
as shown by Fig.~\ref{fig:09a}. 
But its existence enables the appearance of a connection between 
hyperbolic points (with one particle at the crest of the wave), 
which generates further chaos. 

Indeed, close to a separatrix, the distances between resonances are very small, 
so that for small perturbation values the system can be driven quickly to the chaotic regime.\cite{LiLi13,GuckHol83} 
In our case, in particular, this means that the dynamics becomes strongly chaotic 
even for small initial perturbations $(z_2, w_2)$ of a coinciding-particles solution.

Even for larger energy, as long as $H < P^2 / 4$, 
the wave intensity cannot vanish during the system evolution. 
Indeed, when $I=0$, the energy reduces to $H = (p_1^2 + p_2^2) / 2 = P^2/4 + w_1^2$. 
Therefore, when the energy remains below the threshold $P^2/4$, 
the wave always keeps a finite intensity, and the motion of particles is constrained by the potential well,
which is modulated smoothly with time and never disappears.  

\subsection{Large scale chaos for $H > P^2/4$}
\label{sec:Chaos}

At energy $H = P^2/4$, a new type of solution appears~: the wave may vanish. 
Then, its phase can undergo a $\pi$-jump, and large scale chaos does prevail, 
in a way similar to what was observed by Menyuk and by del-Castillo-Negrete and Firpo \cite{Menyuk,CaFi02}. 

\begin{figure}[!tb]
  \centering
   \subfigure[]{\includegraphics[width=0.40\textwidth]{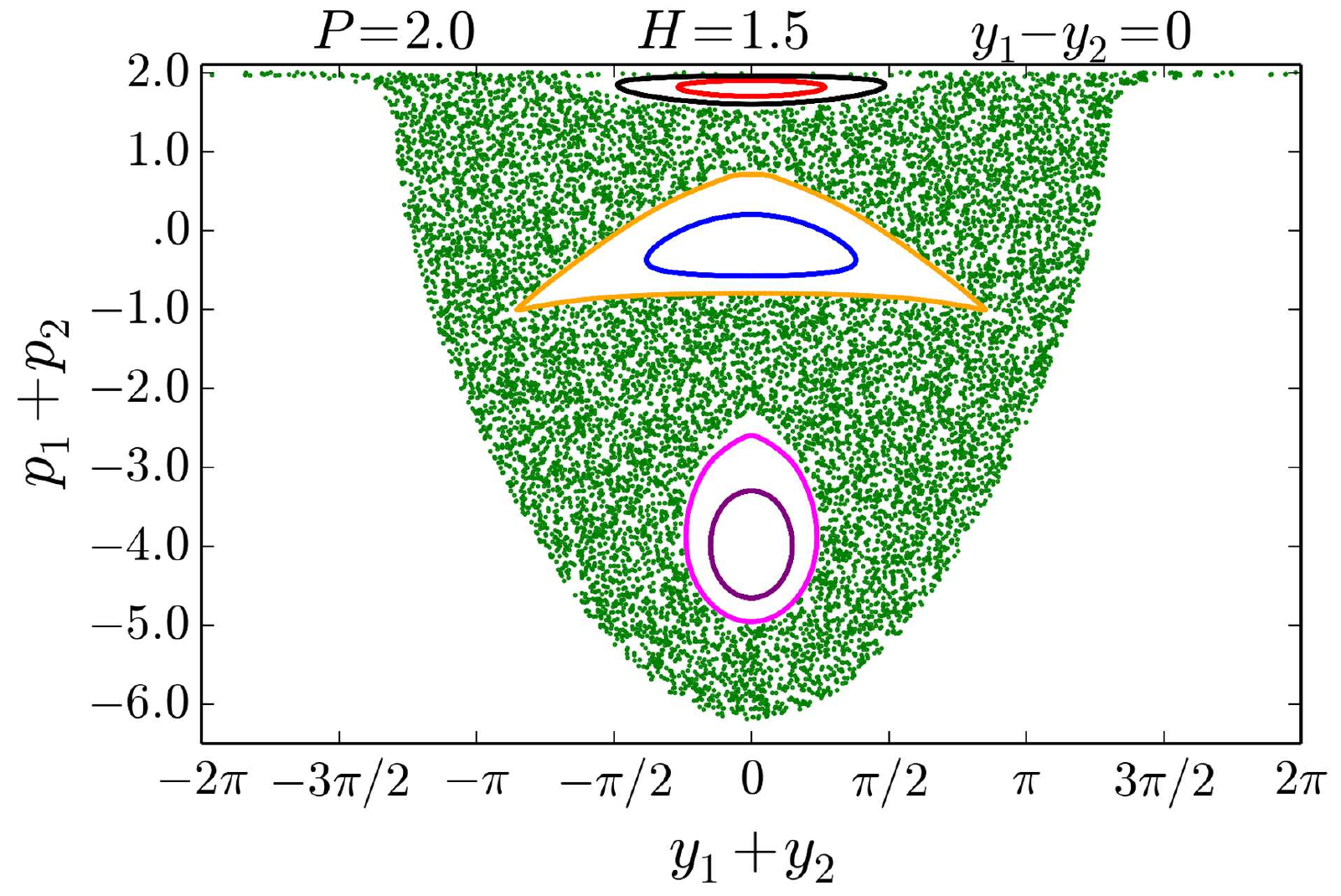}\label{fig:10a} } \\
   \subfigure[]{\includegraphics[width=0.40\textwidth]{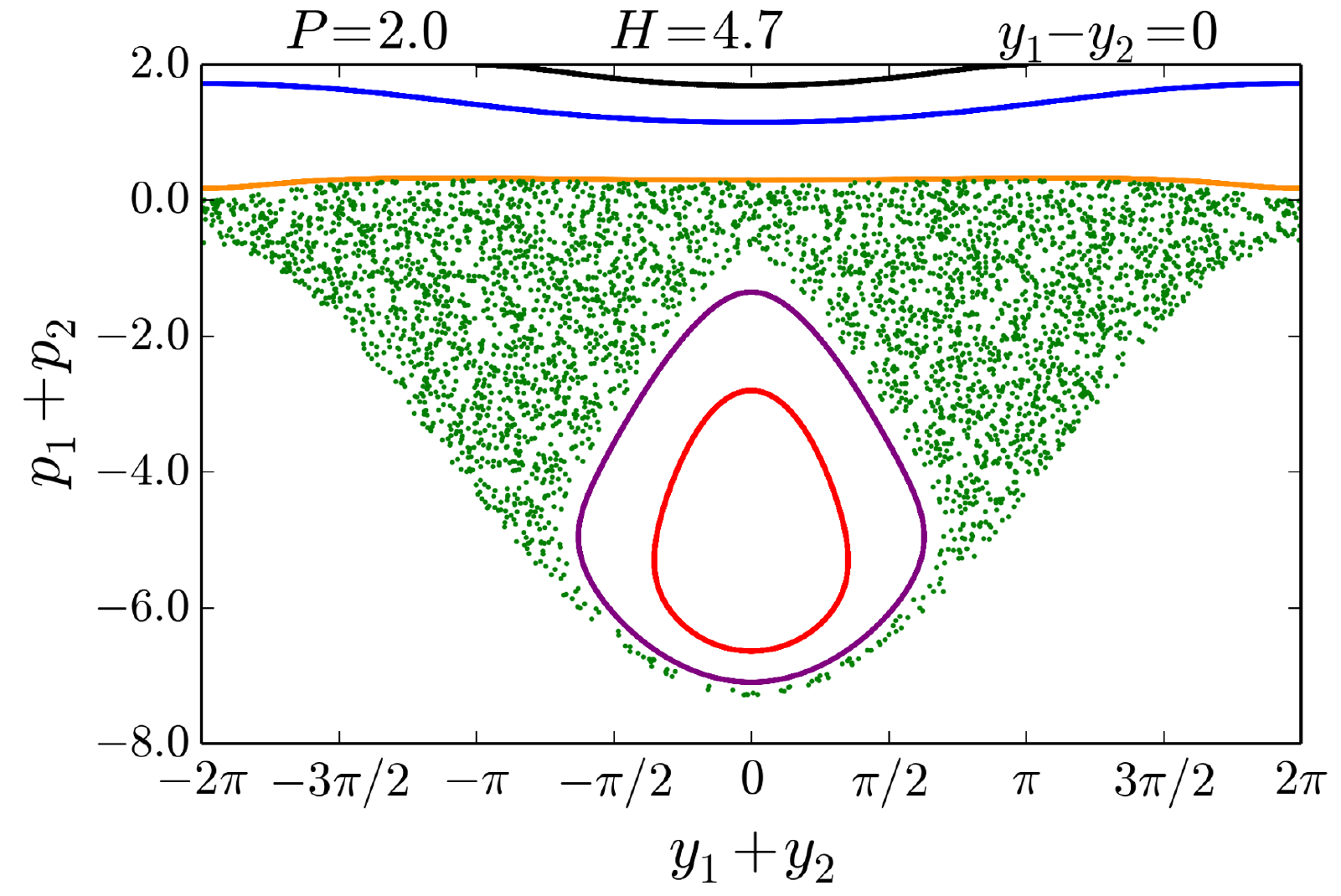}\label{fig:10b} }
   \caption{Interception of trajectories with the Poincar\'e section located at $y_1 - y_2 = 0$ for large positive $H$ values. 
   The system total momentum is $P=2$ on both panels 
   and the total energy increases from (a)~$H=1.5$ to (b)~$H=4.7$.}
   \label{fig:10}
\end{figure}

Poincar\'e sections for larger $H$ values are shown in Fig.~\ref{fig:10}. 
In this energy regime, both particles can wander far away from the bottom of the wave potential well.
In particular, for $H = 1.5$, the green (chaotic) trajectory reaches close to $p_1 + p_2 = P$ (viz.\ $I = 0$), 
allowing a very wide range of values for $y_1 = y_2$. 

\begin{figure}[!tb]
  \centering{
     \includegraphics[width=0.40\textwidth]{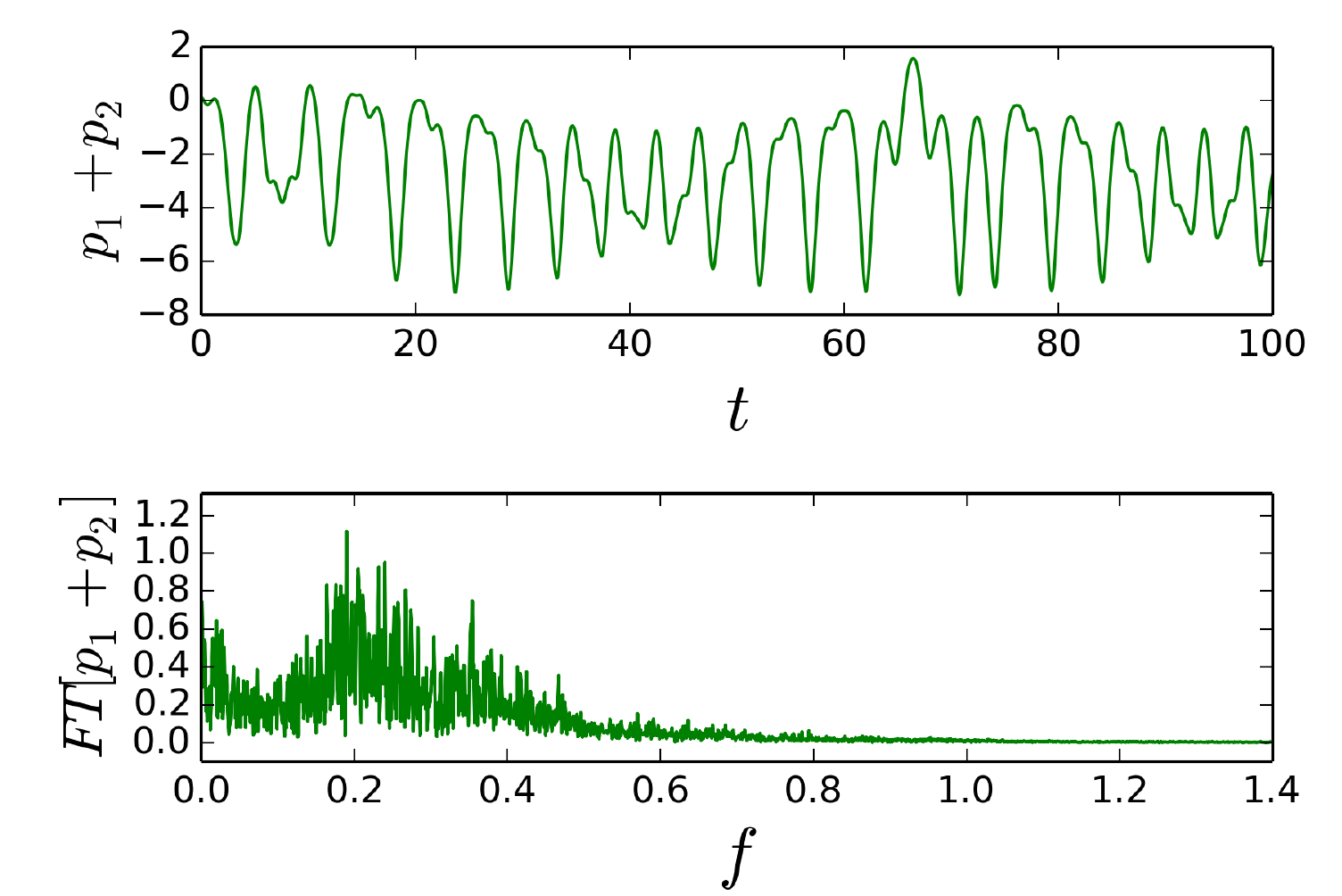}
   \caption{Time evolution and Fourier transform of the particles total momentum 
   for the green chaotic trajectory in Fig.~\ref{fig:10b}.}
   \label{fig:11}}
\end{figure}

The time evolution of the particles total momentum and its Fourier transform 
for the green chaotic trajectory in Fig.~\ref{fig:10b} is shown in Fig.~\ref{fig:11}. 
The Fourier transform looks just like a noise 
and it is not possible to point out a peak frequency. 
This may suggest that, in this chaotic regime, 
particles are free to move in phase space after gaining energy from the wave. 
The particles come back to exchange energy with the wave 
because the system is conservative and motions occur on compact manifolds. 

\section{Conclusion}
\label{sec:Conclusion}

In this work, we analyze the regular and chaotic dynamics in the wave-particle interaction 
using the self-consistent Hamiltonian model.\cite{ElEs03,EsEl03} 
Considering the single wave model,\cite{OLM70,OWM71} we study the dynamics for $N = 1$ and $N = 2$ particles. 

In the first stage, we recall the analysis\cite{TenMeMo94,Adam1981,Ca02} 
of the self-consistent wave-particle interaction for $N = M = 1$. 
As this system is integrable, its phase space presents only regular trajectories. 
Integrable cases are important because they provide~\cite{Pakter95} a basic understanding of the coherent structures found for large $N$. 
As observed in subsection~\ref{sec:PhasePortraitM1N1} for different values of total momentum, the phase portrait topology of the $H=$ constant contours changes. 
For the specific value $P=3/2$, the system has a bifurcation point at which an elliptic-hyperbolic pair of fixed points coalesce. 

Bifurcation diagrams, Figs~\ref{fig:BifDiag} and \ref{fig:PxH}, provide a clear description of the system dynamics in terms of equilibrium solutions. The analysis of the phase portrait complements the bifurcation diagrams. 
After the saddle-center bifurcation, a separatrix orbit appears and divides the phase portrait topology in three different domains, and the evolution of the system is different in each domain.
Moreover, 
for the special value $P=3/4^{1/3}$, the system has a global bifurcation 
by which the energy line that contains $I=0$ passes through the hyperbolic point.

For $N = 2$, we identify and analyze the emergence of chaos in a low-dimensional system. 
In this scenario, the discussion about the chaotic activity can be divided into two regions of phase space, 
namely: close to the hyperbolic and elliptic fixed points. 

The appearance and intensification of chaos in the region close to the hyperbolic fixed point is usual, since, for $N = 2$ the system is non-integrable, and the homoclinic tangle generated from a separatrix in a non-integrable Hamiltonian system is a skeleton near which chaotic transport develops. 
Chaos in this scenario is called  separatrix chaos. 
In this region of phase space, the system presents strong sensitivity on the initial condition, 
so that the interaction quickly leads to chaos for small variations in the particles relative position and velocity. 

On the other hand, for wave-particle systems, the appearance of chaos near the elliptic fixed point is not typically expected. 
For negative $H$, the momenta $p_r$ associated with the particles are small, 
whereas the wave intensity $I$ is large since total momentum $P$ and total energy $H$ 
are conserved quantities for this dynamics. 
Therefore, in the beginning, the particles should move in the wave potential well and, 
as we increase the disturbance in the system, 
the particles would have more energy to exchange with the wave, 
while remaining trapped. 
Our results show that the contribution of the resonance is eventually enough 
to destroy tori and establish chaos in this domain. 
A more appropriate view of the $N=2, M=1$ system in this regime is that 
of two particles coupled through an effective interaction mediated by the wave,  
similar to the low-energy regime of the celebrated H\'enon-Heiles system \cite{HeHe64,Low12}. 
Finally, for large enough energy, the wave intensity can occasionally vanish, which results in very efficient chaos. 

The description of the wave-particle interaction in low-dimensional approximation 
proved to be effective in analyzing basic characteristics of the system, 
related to the emergence and intensification of chaos for $N = 2$. 
Similarly, investigating the ``mirror'' case of $N=1$ particle coupled to several waves 
is also an important issue, which will be discussed in a separate work. 
However, while cases $N + M = 3$ can be thoroughly investigated, 
increasing the number of degrees of freedom toward the $ N \gg 1$ case 
(and similarly $M \gg 1$) 
is the real challenge for a sharp understanding of the fundamental problem 
of the transition from dynamics to statistical behavior.~\cite{ElEs03, EsEl03,ElFi98}

\begin{acknowledgments}
The authors acknowledge discussions 
with Drs C.~Chandre, X.~Leoncini, L.~H.~Miranda F., T.~M.~Rocha Filho, 
and members of the \'equipe turbulence plasma in Marseille. 
They also thank anonymous reviewers for constructive comments. 

The Centre de Calcul Intensif d'Aix-Marseille is acknowledged 
for granting access to its high performance computing resources. 
JVG thanks Coordena\c{c}\~ao de Aperfei\c{c}oamento de Pessoal de N{\'\i}vel Superior (CAPES) for financing her stay at Aix-Marseille Universit\'{e} (AMU) under the Programa de Doutorado Sandu{\'\i}che no Exterior (PDSE), process No. 88887.307528/2018-00, and Conselho Nacional de Desenvolvimento Cient\'ifico e Tecnol\'ogico (CNPq) for a doctoral fellowship at Universidade Federal  do Paran\'a (UFPR), process No. 166914/2017-7.
MCS thanks CAPES for financing her stay at AMU under the Programa Est\'agio P\'os-Doutoral no Exterior, process No. 88887.307684/2018-00, 
and the Funda\c{c}\~ao de Amparo \`a Pesquisa do Estado de S\~ao Paulo (FAPESP) for a postdoctoral fellowship at Universidade de S\~ao Paulo (USP) under grant No. 2015/05186-0 (associated with grant No. 2018/03211-6). At the beginning of this work, 
YE enjoyed the hospitality of the grupo controle de oscila\c{c}\~oes at USP, and RLV and ILC enjoyed the hospitality of the \'equipe turbulence plasma at AMU, with the support from a COFECUB-CAPES grant Nos. 40273QA--Ph908/18 (COFECUB - Comit\'e Fran\c{c}ais d'\'Evaluation de la Coop\'eration Universitaire et Scientifique avec le Br\'esil), and 88881.143103/2017-01 (CAPES). 
RLV received financial support from CNPq, process No. 301019/2019-3.
ILC acknowledges financial support from FAPESP under grant No. 2018/03211-6, and CNPq under grant Nos. 407299/2018-1 and 302665/2017-0.
\end{acknowledgments}

\section*{Data availability}
 The data that support the findings of this study are available from the corresponding authors upon reasonable request. 

\nocite{*}
\bibliography{Gomes_2021_wave_particle}

\end{document}